# Electron spin dynamics guide cell motility

**Authors**: Kai Wang,[1-3]* Gabrielle Gilmer,[1,2,4,5]* Matheus Cândia Araña,[1,2,6] Hirotaka Iijima,[1-3] Juliana Bergmann,[1,2] Antonio Woollard,[7] Boris Mesits,[7] Meghan McGraw,[1,2] Brian Zoltowski,[8] Paola Cappellaro,[9,10] Alex Ungar,[9,10] David Pekker,[7] David H. Waldeck,[11] Sunil Saxena,[11] Seth Lloyd,[12] Fabrisia Ambrosio[1-3]**

**Affiliations**:

[1]Discovery Center for Musculoskeletal Recovery, Schoen Adams Research Institute at Spaulding, Boston, MA
[2]Department of Physical Medicine & Rehabilitation, Spaulding Rehabilitation Hospital, Boston, MA
[3]Department of Physical Medicine & Rehabilitation, Harvard Medical School, Boston, MA
[4]Cellular and Molecular Pathology Graduate Program, University of Pittsburgh, Pittsburgh, PA
[5]Medical Scientist Training Program, School of Medicine, University of Pittsburgh, Pittsburgh, PA
[6]École Polytechnique, Institut Polytechnique de Paris, Palaiseau, France
[7]Department of Physics & Astronomy, University of Pittsburgh, Pittsburgh, PA
[8]Department of Chemistry, Southern Methodist University, Dallas, TX
[9]Department of Nuclear Science and Engineering, Massachusetts Institute of Technology, Cambridge, Massachusetts 02139, United States
[10]Department of Physics, Massachusetts Institute of Technology, Cambridge, Massachusetts 02139, United States
[11]Department of Chemistry, University of Pittsburgh, Pittsburgh, PA
[12]Department of Mechanical Engineering, Massachusetts Institute of Technology, Boston, MA
* Equal contribution

**Corresponding Author:

Fabrisia Ambrosio, PhD, MPT

Suite 5.303, 149 13th St, Charlestown, MA 02129

Phone: (412) 657-1525

Email: fambrosio@mgh.harvard.edu

**Abstract**

Diverse organisms exploit the geomagnetic field (GMF) for migration. Migrating birds employ an intrinsically quantum mechanical mechanism for detecting the geomagnetic field: absorption of a blue photon generates a radical pair whose two electrons precess at different rates in the magnetic field, thereby sensitizing cells to the direction of the GMF. In this work, using an *in vitro* injury model, we discovered a quantum-based mechanism of cellular migration. Specifically, we show that migrating cells detect the GMF via an optically activated, electron spin-based mechanism. Cell injury provokes acute emission of blue photons, and these photons sensitize muscle progenitor cells to the magnetic field. We show that the magnetosensitivity of muscle progenitor cells is (a) activated by blue light, but not by green or red light, and (b) disrupted by the application of an oscillatory field at the frequency corresponding to the energy of the electron-spin/magnetic field interaction. A comprehensive analysis of protein expression reveals that the ability of blue photons to promote cell motility is mediated by activation of calmodulin calcium sensors. Collectively, these data suggest that cells possess a light-dependent magnetic compass driven by electron spin dynamics.

**Introduction**

Electron reorganization is central to many chemical processes that govern biological systems, including energy transfer, chemical bonding, and molecular signaling. Electrons have two fundamental quantum properties, charge and spin. The importance of electron charge has been widely investigated for cellular processes such as ion transport, enzyme catalysis, and oxidative phosphorylation. In contrast, electron spin, which reflects an electron's angular momentum, remains largely unexplored in biology. Understanding how spin effects operate at the cellular level could deepen our comprehension of quantum-scale biological processes, shedding light on fundamental aspects of cellular behavior.

Electron spin is intrinsically linked to magnetic properties because the spin of a charged particle generates a magnetic moment that allows the electron to interact with external magnetic fields, such as the geomagnetic field (GMF). Among the most compelling evidence of spin-related biological responses is the proposed mechanism for the avian migratory compass.(*1, 2*) In this proposed, mechanism, absorption of blue light from the sun, moon, or stars initiates an electron transfer reaction that generates a radical pair in photosensitive proteins present within the retina.(*3*) Radical pairs are chemical intermediates comprising two electrons that are localized on different regions of a molecule and are weakly coupled to one another. These unpaired electrons have their spins arranged in either a singlet state (spins anti-aligned with a total momentum of 0) or a triplet state (spins either aligned or anti-aligned but with a total momentum of 1). Transitions between these spin states can be influenced by magnetic fields, such as the GMF. Whether the radical pair is triplet or singlet can impact the subsequent chemistry of the radicals, including their recombination rates and the chemical identity of the reaction products.(*4, 5*) These findings shed light on the role of magnetic sensitivity in animal navigation. However, the broader implications of electron spin dynamics for cellular processes remain largely unexplored.

This paper establishes the presence of a cell magnetic compass that is governed by electron spin dynamics. We focused on muscle progenitor cells (MPCs) because skeletal muscle repair requires the direction-dependent migration of these resident cells to the injury site, where they differentiate and fuse into newly regenerated fibers.(*6*) Previous work has shown magnetic fields on the order of 80 mT promote myoblast differentiation and fusion(*7, 8*), effects typically attributed to classical physics cellular responses. However, it is unclear whether MPCs are sensitive to ultraweak magnetic fields such as the GMF, a question that could uncover a quantum-based mechanism of cellular sensitivity. Using 2D and 3D *in vitro* models of cell injury, we demonstrated that: (1) **Endogenous blue photons promote MPC motility**. Cell injury generates acute blue photon emission, and quenching of these photons inhibits motility. (2) **Blue photon-induced motility is dependent on the GMF.** The ability of blue photons to promote motility relies on the presence of the GMF; (3) **Electron spin dynamics govern sensitivity to the GMF**. Disruption of coherent electron spin precession via the application of an oscillatory field at the free electron Larmor frequency diminishes the influence of the GMF on MPC motility.

Given the sparsity of photoreceptors and minimal light penetration beyond the skin and eyes, our findings challenge the conventional wisdom that internal cells and tissues are unlikely to be photosensitive and implicate quantum coherent reactions in cellular responses. Our results show that MPCs possess a quantum compass, suggesting that nature may have evolved similar mechanisms across various biological contexts. We anticipate that the light/GMF interactions observed here will extend to a broad range of phenomena.

## Results

*Muscle cells are sensitive to magnetic fields in the microtesla range*

Previous work demonstrated that elimination of the GMF inhibited motility of neuroblastoma cells.(*9*) To test whether MPCs are similarly responsive to cues emanating from the GMF, we utilized a wound scratch model by injuring a monolayer of MPCs *in vitro*.(*10*) Injured monolayers were then cultured either in a Mu-metal Faraday cage to shield cells from the GMF or under normal culture conditions as a control (**Figure 1A, B**). We confirmed that the GMF decreased from 48-52 µT to ~5 nT in the Faraday cage with no temperature difference across the two environments. By measuring scratch distance, whereby a decrease in distance indicates increased migration, we found elimination of the GMF significantly decreased MPC migration (**Figure 1C, D**).

The GMF corresponds to energy levels $10^7$ times lower than the available thermal energy, suggesting MPC sensitivity to the GMF arises from spin transitions that are weakly coupled to the thermal bath. In support of a quantum-based mechanism for avian migration, application of a radiofrequency field at the Larmor frequency of an electron in the GMF (1.35 MHz) disorients birds.(*11-13*) This effect is expected to arise from resonance of the applied radiofrequency field with the electron's spin precession, which enhances the mixing of singlet and triplet pair states. We therefore wanted to know whether radiofrequency fields can disrupt cell migration in the GMF (**Figure 1E**). We found that wound closure was significantly decreased with application of a 1.35 MHz radiofrequency field, but was not affected by the oscillating field at 11.2 MHz, a frequency that does not resonate with electron spin precession under the GMF (**Figure 1F**). These data suggest that MPCs sensitivity to the GMF relies on electron spin dynamics.

Cell migration for wound healing is a highly direction-dependent process, and we hypothesized that MPC magnetosensitivity serves to orient cells at a site of injury. We therefore evaluated whether MPCs are sensitive to magnetic field direction. Scratched cell monolayers were exposed to ultraweak magnetic fields (400 µT) oriented along the *X* or *Y* direction in a magnetic field chamber (**Figure 1G, H**). We chose 400 µT because previous work showed that fields in this range stimulate regeneration in a planarian amputation model.(*14*) Our data showed strong sensitivity of cells to magnetic field directions (**Figure 1I**). To further demonstrate MPC magnetosensitivity and its relevance to muscle functional regeneration, we used a three-dimensional (3D) bioengineered muscle construct model and evaluated muscle regeneration and MPC behavior under magnetic field stimulation.(*15*) Injured muscle constructs were exposed to a 400 µT magnetic field aligned in a direction parallel to the myofibers (**Figure 1J**). Evaluation of MPC lineage progression one day after injury revealed that the percentage of activated (Pax7+/MyoD+), and differentiating (Pax7-/MyoD+) MPCs was increased following magnetic field exposure, while the number of quiescent MPCs (Pax7+/MyoD-) remained similarly low in both groups. (**Figure 1L, M**). Consistent with enhanced MPC activation, contractile function was significantly increased seven consecutive days of magnetic field stimulation when compared to non-stimulated construct controls (**Figure 1K**). Collectively, our data suggest that ultraweak magnetic fields contribute to MPC responses critical for regeneration.

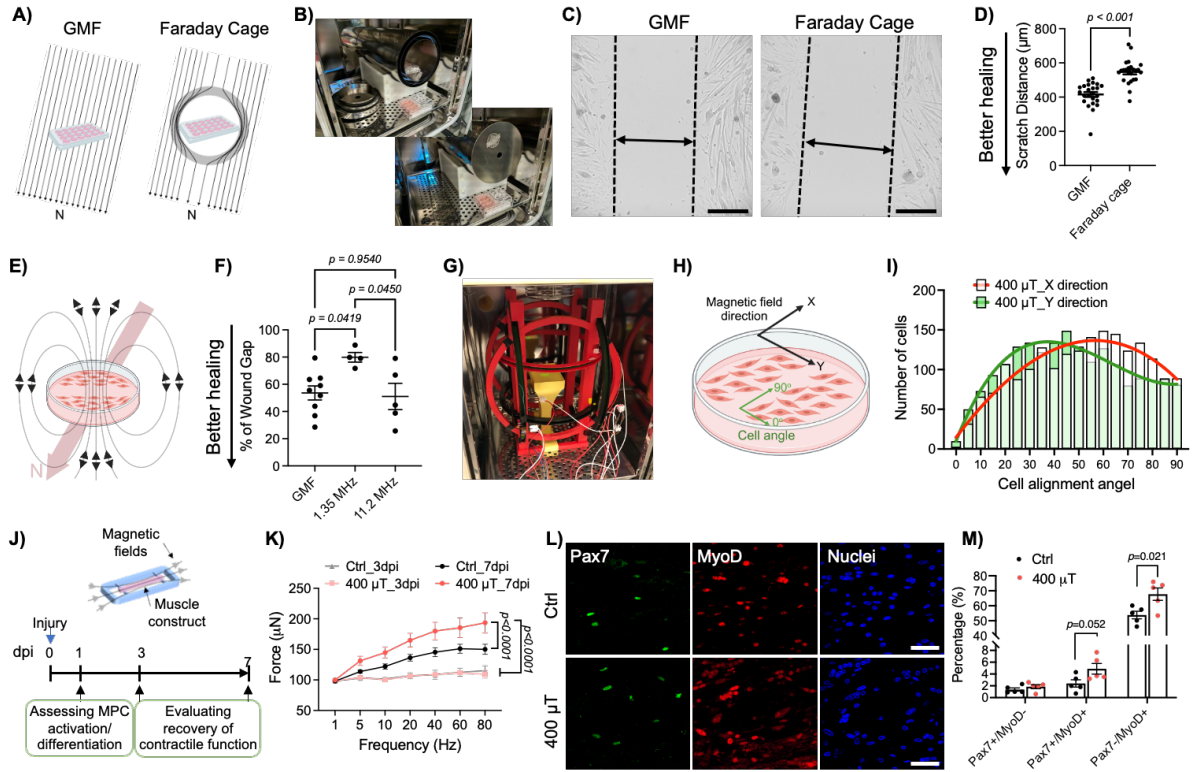

**Figure 1. MPCs are sensitive to weak magnetic fields in both 2D and 3D systems.** (A) Schematic of cells under the geomagnetic field (GMF) and in a Mu-metal Faraday cage. (B) Photographs of scratched cells in the Mu-metal Faraday cage placed inside a cell culture incubator. (C) Representative wound scratch images and (D) scratch distance of cells under each condition for 6 hours. Dashed lines highlight wound areas and double arrow lines indicate wound distances. Scale bar: 300 µm. Two-tailed student's *t*-test, n=24/group. (E) Schematic of scratched cells in the presence of an oscillating field. The Earth's magnetic field is illustrated by the thick pink arrow. (F) Percentage of the remnant wound gap after 6 hours of 1.35 MHz or 11.2 MHz oscillating field stimulation. Scratched monolayers that were not stimulated with oscillating fields were used as controls (GMF). One-way ANOVA. n=4-9/group. G) A photograph of the magnetic field stimulation setup. (H) Schematic illustration of magnetic field direction and cell alignment angle. (I) Histogram of cell angle distribution when cells were stimulated in the *X*- versus Y-direction. (J) Schematic illustration showing the magnetic field stimulation of a 3D muscle construct. dpi: days post injury (K) Force-frequency curves of injured muscle constructs at 3 and 7 dpi. Two-way ANOVA using a mixed model. n=4/group (3 dpi), n=12 (7 dpi). (L) Representative confocal images of injured muscle constructs stained for Pax7 (green) and MyoD (red) at 1 dpi. Nuclei were stained by DAPI in blue. Scale bar: 50 µm. (M) Quantification of Pax7+/MyoD-, Pax7+/MyoD+, or Pax7-/MyoD+ nuclei in the injured constructs one day post injury. Two-tailed Student's t-test, n= 4/group

*Cell injury provokes localized blue photon emission*

In the avian compass, radical pairs are hypothesized to be generated by photons from the sun, moon, and stars. Extrapolating from this concept, we hypothesized that absorption of photons by MPCs initiates an analogous reaction that sensitizes cells to the GMF. This led us to a pivotal question: what might be the source of such photons under conditions of tissue injury? Previous work in a plant model showed that acute tissue damage induces increased endogenous photon emission.(*16*) This prompted us to investigate whether injury to a myoblast monolayer induces photon emission and, if so, to characterize the properties of these emitted photons. Thus, we recorded photon emission of myoblast monolayers at baseline (uninjured), immediately after injury (within one hour), and 24 hours after injury (**Figure 2A, S1A**). Photon emission was detected in uninjured cells (**Figure S1A**), consistent with previous work demonstrating that cells under homeostasis conditions emit photons. We confirmed that photons originated from the cells and not the culture dish or media (**Figure 2B, S1A**). Immediately after injury, photon emission increased by 1.46 ± 0.28-fold compared to baseline (**Figure 2C).** Photon emission was further increased by 1.84 ± 0.36-fold 24 hours after injury. To estimate the wavelength range of emitted photons, we recorded photons with bandpass filters for blue (430-470 nm), green (530-570 nm), and red (630-670 nm) light. Blue photon emission increased by 2.5 ± 0.1-fold immediately following injury, returning to baseline by 24 hours after injury (**Figure 2D**). In contrast, green photon emission displayed minimal changes immediately after injury, but a 1.79 ± 0.07-fold increase 24 hours after injury (**Figure 2E**). Red photon emission changed minimally before and after injury (**Figure 2F**).

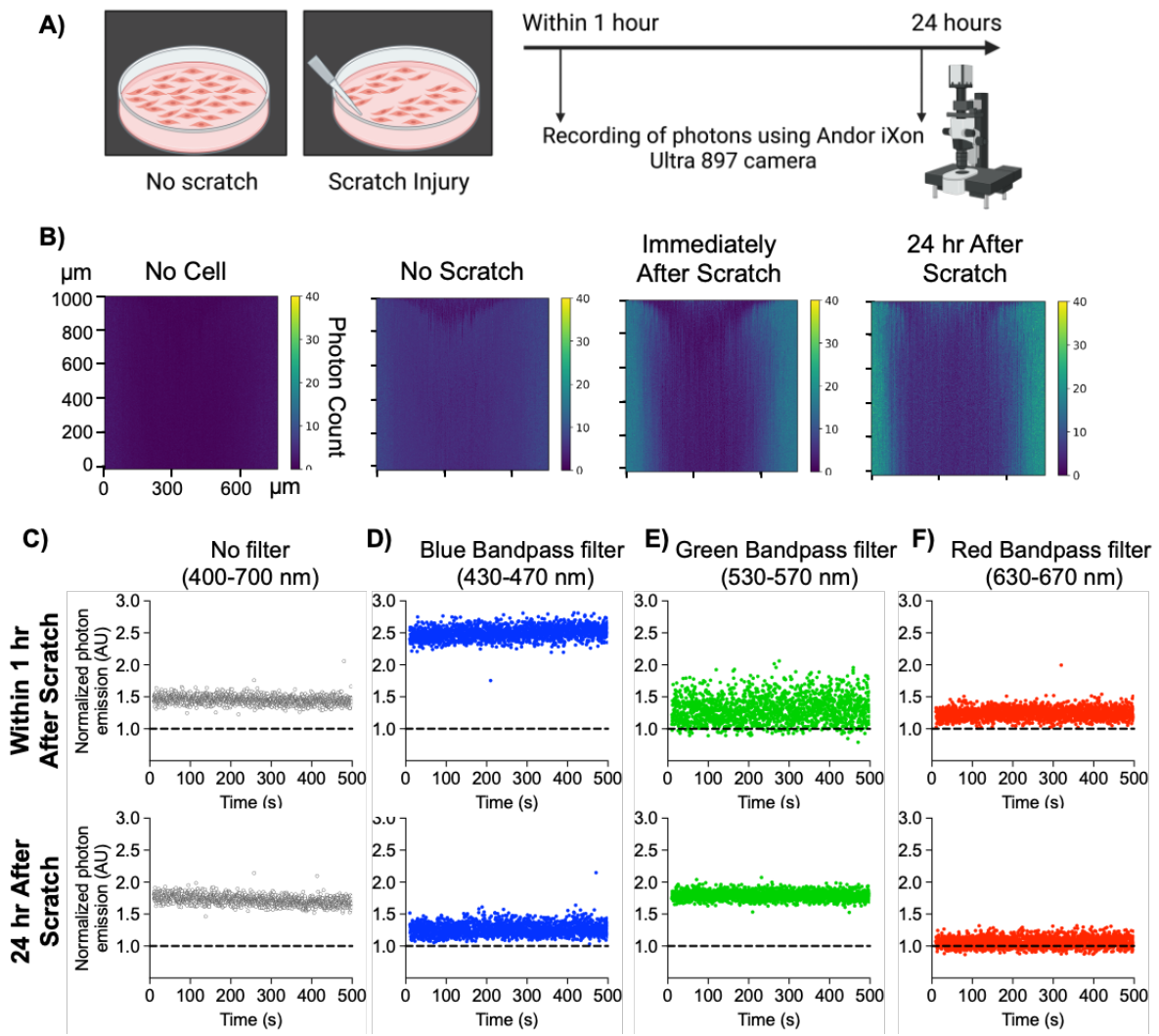

**Figure 2. Scratch wounds of myoblasts emit photons *in vitro*.** A) Illustration of photon recording time points. B) Representative photon emission of cell culture dishes (no cells), cells without scratch, cells immediately after scratch, and cells 24 hours after scratch. The photon emission was collected over a 500-second window. C-F) Normalized photon emission of cells immediately after scratch and 24 hours after scratch. Photon count per cell count in injured groups was normalized to the non-injured cell run on the same day, with a value of 1 representing the non-injured sample normalized to itself (black dashed lines). Bandpass filters were used to record the emission of blue, green, and red photons. n=2-3.

*Light stimulation enhances MPC migration in a time- and intensity-dependent manner*

The above data imply that acute injury increases photon emission, with blue photons dominating immediately after injury. Given the presence of natural bioluminescence in cells, we hypothesized that photon emission promotes wound healing, and we evaluated the effect of irradiation on wound healing and MPC motility. We first tested whether MPCs respond to photon stimulation. Scratched cell monolayers were randomized to one of four experimental groups consisting of 0, 0.1, 1, or 10 second(s) of white light exposure per hour at an intensity of 6,844 mW/m$^2$ for 6 hours (**Figure S1C-D**). We found that scratch distance decreased following 1 and 10 seconds of light exposure per hour compared to controls (**Figure 3A**). As a secondary metric of cell migration, we quantified F-actin expression by immunofluorescence staining. A lower F-actin fluorescent level is associated with faster and more recent cell migration.(17) Consistent with decreased scratch distance, F-actin levels were significantly lower following photon-stimulation (**Figure 3B**). F-actin intensity was highly correlated with scratch distance (**Figure 3C**), further suggesting that the enhanced healing is a function of increased cell migration. We excluded the possibility of thermal effects by confirming that the temperature remained constant over a ten-minute exposure at the maximum light intensity (**Figure S1B**). Given that 1 second of light stimulation per hour was sufficient to enhance wound closure and cell migration, all subsequent experiments utilized this protocol.

Next, we evaluated the effect of light intensity on cell migration. Scratched cell monolayers were randomized to light exposure at different LED power settings for 6 hours: 6,844 mW/m² irradiance, 287 mW/m², 78 mW/m², 30 mW/m², 21 mW/m², 15 mW/m², or in the dark. Exposure of cells to light at 21 mW/m² resulted in the greatest wound closure and migration compared to other exposure groups, with increased intensities suppressing the effect (**Figure 3D-E**).

*MPC migratory responses to light are blue-photon specific and affected by the GMF*

We next sought to determine whether the observed light stimulation effects were wavelength-dependent. For this, another cohort of injured monolayers were randomized and irradiated with blue (430-470 nm), green (530-570 nm), or red (630-670 nm) light (**Figure 3F**). Cells exposed to either white light or dark conditions served as positive and negative controls, respectively. Neither red nor green light exposures affected wound closure or migration markers (**Figure 3G-H**). However, white light responses were recapitulated when cells were exposed to blue light (**Figure 3G-H**). We confirmed that blue photon-enhanced cell migration is not attributed to cell proliferation, apoptosis, and necrosis (**Figure S2B**) and that the cell responses to light were not secondary light-induced alterations in the media (**Figure S2C, D**). We also repeated the above experiments using young female muscle stem cells, and for all metrics collected, we obtained results similar to those observed in male cells (**Figures S3**). Notably, the beneficial effect of blue light stimulation on MPC migration was abrogated when cells were cultured in a Mu-metal Faraday cage (**Figure 3I, J**), suggesting the GMF is required for photo-induced cell motility responses. In a loss-of-function paradigm, we added a blue photon absorbing molecule, chlorophyll b, to the cell media following wound injury (**Figure 2K**). These studies were performed in the dark so that chlorophyll b would absorb endogenously emitted photons. We confirmed that neither cell morphology nor viability were affected by chlorophyll b (**Figure S2A**). The presence of chlorophyll b significantly attenuated wound closure **(Figure 2L)**. Taken together, these data support the conclusion that endogenous blue photons increase cell migration in a manner that may be dependent on the GMF.

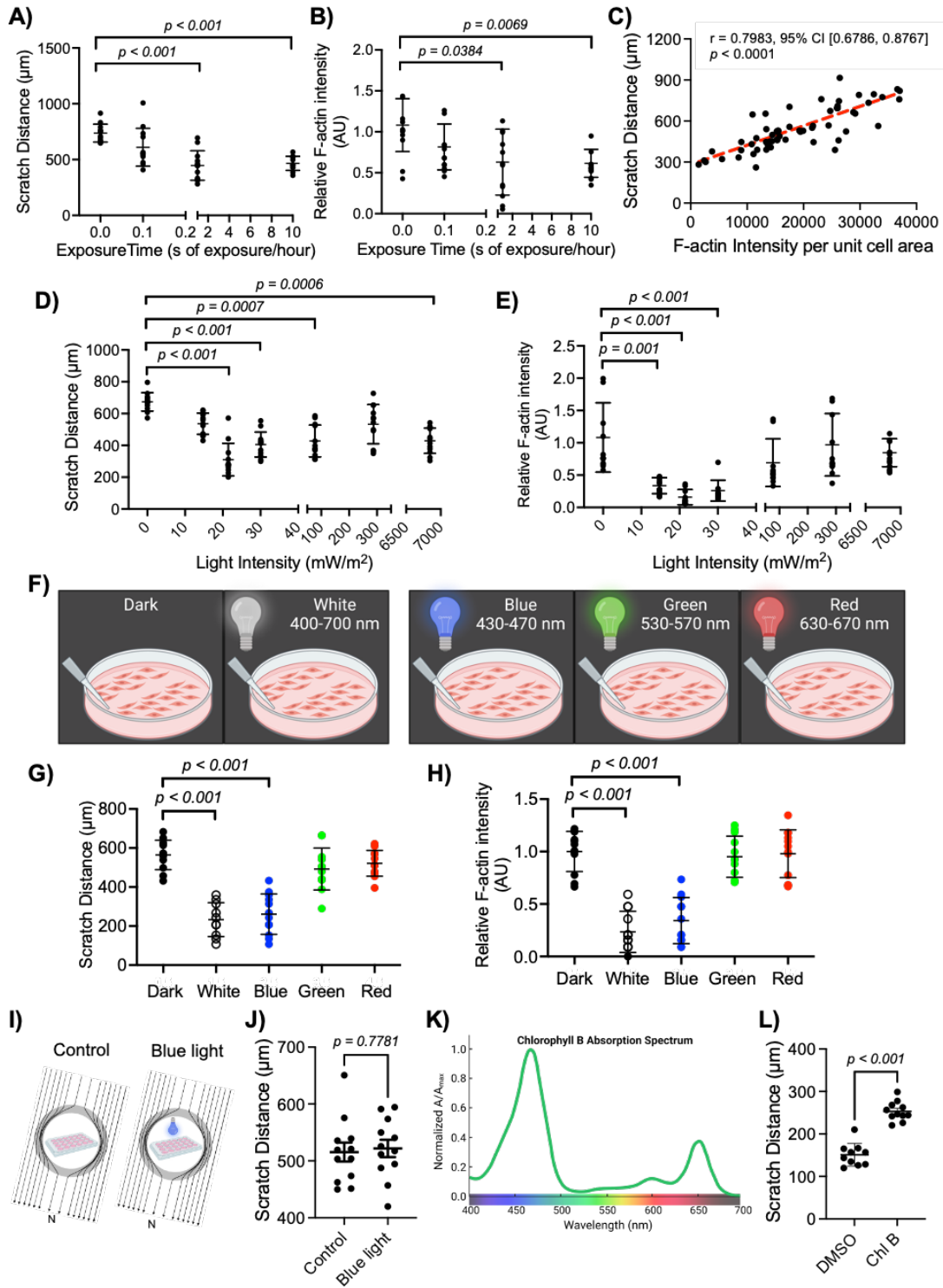

**Figure 3. MPC migratory responses are specific to blue photons.** A-B) Scratch distance and relative F-actin intensity of scratched cells after 6 hours of light exposure at various light exposure times. Kruskal Wallis tests, n=12/group. C) Correlation between F-actin intensity and Scratch distance. The *p* value was calculated using a Pearson correlation test. D-E) Scratch distance and relative F-actin intensity of scratched cells after 6 hours of light exposure at different light intensities. Kruskal Wallis tests, n=12/group. F) Experiment design. G-H) Scratch distance and relative F-actin intensity versus light spectra. One-way ANOVA, n=10-12/group. I) Schematic illustration of blue light stimulation inside the Mu-metal Faraday cage. J) Wound distance of scratched cells after 6 hours of blue light stimulation. In the control group, scratched cells were placed in the Mu-metal Faraday cage for 6 hours and covered by aluminum foil. Two-tailed student's *t*-test, n=12/group. K) Absorption spectrum of chlorophyll b. L) Scratch distance of scratched cells after 6 hours of chlorophyll b treatment (10 ⌈M; Chl B). Same amount of DMSO was added to the cells in the control group.  Two-tailed student's *t*-test, n=10/group.

*Blue photons stimulate calmodulin proteins and protein modifications associated with motility*

In migratory birds, the photosensitive protein, Cryptochrome, in the retina is hypothesized to initiate a radical pair reaction upon blue-green light stimulation.(*18*) To test a possible role of Cryptochrome in modulating the observed light-based effects on MPC migration, we used siRNAs to knockdown the two Cryptochromes found in mammalian cells, Cryptochrome 1 (Cry1) and Cryptochrome 2 (Cry2) (**Figure S4A-C**). We also knocked down period 1 (Per1) and period 2 (Per2), two proteins with circadian functions, to control for dysregulation of circadian patterns in cells.(*19*) Knockdown of these proteins did not affect MPC responses to blue light (**Figure S4C-E**), suggesting that Cry1 or Cry2 may not be the mediator of blue light effects on MPC motility.

To interrogate the molecular mechanisms by which blue photons enhance MPC migration, we performed multi-omics analyses using RNA-sequencing and mass spectrometry-based proteomics. Transcriptomic data contained 9,947 unique genes and proteomic data contained 7,704 unique proteins. However, MPC responses to blue photon stimulation were heavily biased towards a shift in protein levels with minimal changes in gene expression. Proteins significantly increased after blue photon stimulation were associated with cell migration-related functions, including "*Regulation of focal adhesion assembly*" and "*Regulation of actomyosin structure organization*" (**Figure 4A**).

Next, we sought to identify signaling pathways involved in blue photon-stimulated cell motility by utilizing a novel computational model, "directional gene set enrichment analysis

(dGSEA).(*20*) The dGSEA uses input gene ranks calculated through correlation analyses between blue photon-stimulated migration proteins (i.e., focal adhesion proteins) and the remaining proteins (target proteins) within the mass spectrometry-based proteomics data set (**Figure 4B**). The dGSEA revealed 41 Reactome pathways that were either positively or negatively associated with the migratory proteins responsive to blue photon stimulation, with calmodulin-related pathways displaying the greatest enrichment (**Figure 4C**). Calmodulin is a well-known intracellular $Ca^{2+}$ sensor and signaling protein that has been linked to a variety of cellular responses, including migration.(*21*) Of the pathways associated with blue photon-stimulated migration, six calmodulin family proteins (Camk2a, Camk2b, Camk2d, Camk2g, Camkk1, and Calm1) were identified as functional hubs (i.e., proteins with the highest degree of overlap between different pathways; **Figure 4D**). To further assess the role of calmodulin in mediating photon-induced cell migration, we performed *in silico* perturbation of those six calmodulin proteins through a network propagation approach. We first constructed a knowledge-based protein interactive network using Stringdb database. We then applied network propagation *via* a random-walk-restart (RWR) algorithm(*22*) to perturb calmodulins on the network, thereby pseudo-activating the corresponding neighboring proteins (**Figure 4E**). Our analyses revealed that activation of the calmodulin pathway resulted in significant and positive enrichment for focal adhesive proteins, recapitulating the MPC responses to blue photon stimulation (**Figure 4F**). Collectively, these analyses suggest that the calmodulin proteins act as mediators for blue photon-induced MPC migration after wound injury.

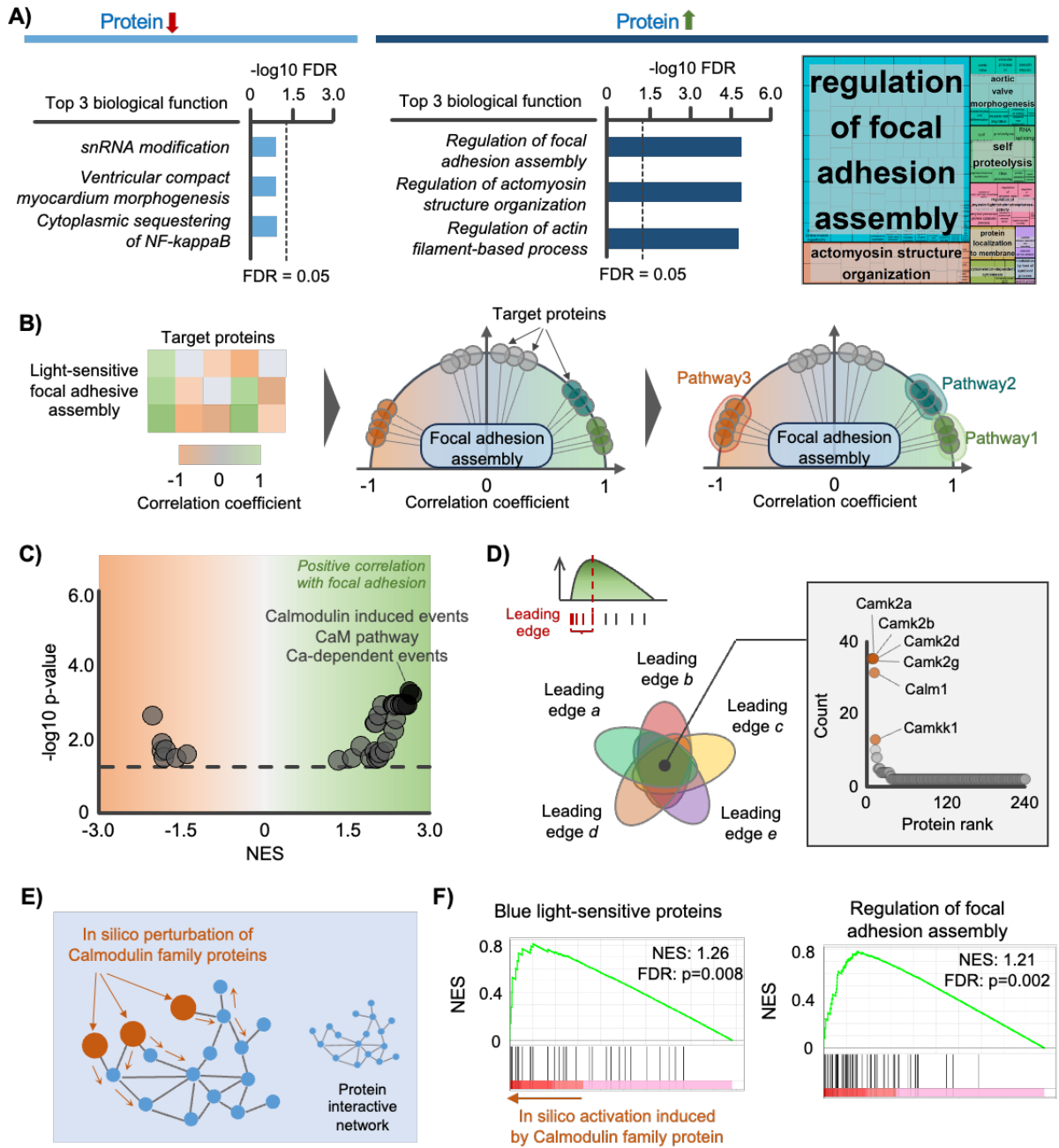

**Figure 4. Calmodulin proteins are associated with blue light-induced MPC motility after injury.** A) Proteins significantly increased after blue light were associated with cell migration-related biological functions, as documented by enrichment of "regulation of focal adhesion assembly" and "actomyosin structure organization". Due to the high redundancy of gene ontology, redundant gene ontology terms were summarized via REVIGO software and visualized as a tree-map. Each rectangle represents a supercluster gene ontology terms, visualized with different colors. The size of the rectangles was adjusted to reflect the p-value of the gene ontology term calculated by top gene ontology (i.e., the larger the rectangle, the more significant the gene ontology term). B) Schematic showing the directional gene set enrichment analysis (dGSEA) concept. The analysis used gene ranks based on the association between blue light-sensitive focal adhesive proteins and other proteins (target proteins) as an input. C)

dGSEA revealed that light-sensitive focal adhesive proteins were co-activated with calmodulin pathways, as evidenced by positive normalized enrichment score (NES). D) Downstream leading-edge analysis of dGSEA identified calmodulin family proteins as primary drivers of Reactome pathways co-regulated with light-sensitive focal adhesive proteins (i.e., proteins with the highest degree of overlap between different Reactome pathways). E) Schematic showing the network model to induce in silico perturbation of calmodulin family proteins identified by the leading-edge analysis. Knowledge-based protein interactive network downloaded through Stringdb was used for the analysis. F) Proteins *in silico* activated by calmodulin family proteins perturbation were significantly enriched to blue-light sensitive proteins (differentially expressed proteins of scratch injury vs injury+blue light) as well as blue light-sensitive focal adhesive proteins.

**Discussion**

In this study, we presented three layers of evidence to support our hypothesis that electron spin dynamics dictate cell motility. First, we discovered that MPCs are sensitive to the direction of ultraweak magnetic fields, but that cell migration is reduced when the GMF is eliminated or disrupted. These findings suggest that the GMF may maintain the coherence or extend the lifetime of quantum spin states critical for directed migration. We next hypothesized that the electron spin states originate from the generation of reactive electron pairs initiated by a biochemical event, such as photostimulation. Consistent with this hypothesis, we found that injury provokes blue photon emission, and that blue photon stimulation enhances cell motility. When endogenous blue photons are quenched, motility is compromised, and when the GMF is eliminated, the effect of blue photons on MPC migration is abrogated. Finally, multi-omics analyses with *in silico* network perturbations revealed that blue light triggers modification of calmodulin proteins that may be associated with migratory pathways.

Our findings suggest an electron spin-based photoactivated mechanism for cell motility. The radical pair model may also explain the observation that myoblast migration and fusion are enhanced by the application of ultra-weak oscillatory fields (1.75 µT, 16Hz)(*23*). Since the electron-spin based mechanism depends only on the orientation of the axis of the field, but not on the N-S direction, a weak and slowly varying oscillatory field should have a similar effect to a weak static field. Moreover, despite energy differences of spin states arising from the GMF being substantially below thermal noise, such effects can cause significant photochemical outcomes. For example, triplet-singlet spin dynamics is believed to play a key role in the biochemical mechanism underlying bird migration.(*3*) Our work did not identify chemical identities and hyperfine interactions that are key to the electron triplet-singlet dynamics and can influence the resonance conditions. Future investigation into these interactions and their potential to alter magnetosensitivity will advance our understanding of quantum effects on cell motility.

Although photon emission from biological materials is well-established,(*24*) it remains unclear how such photons are generated. One working theory for photon emission in biological systems is through the relaxation of chemically generated excited electronic states.(*25*) Specifically, when $O_2$ is converted to $O_2^-$, $H_2O_2$, $HO^.$, or $^1O_2$ in cells, triplet excited carbonyls ($^3R=O^*$), singlet excited pigments ($^1P^*$), triplet excited pigments ($^3P^*$), and $^1O_2$, respectively, are generated and emit photons at various spectral wavelengths.(*26, 27*) Previous studies have shown that $^3R=O^*$ emits photons in the near ultraviolet and blue-green spectra (350-550 nm), $^1P^*$ emits in the green-red spectra (550-750 nm).(*26, 27*) Our findings raise the novel hypothesis that injured muscle cells emit blue/green photons via $^3R=O^*$ and/or $^1P^*$.

In retinal cells, three mechanisms of photodetection have been identified: photomechanical, photothermal, and photochemical.(*28-30*) Photomechanical mechanisms are largely driven by high intensity (megawatts to terawatts per cm$^2$), short exposure (picoseconds to nanoseconds) light in a manner independent of light spectra. However, the intensity of light required to enhance healing in our model system is very small (i.e., a few watts per square meter applied for only one second per hour), and responses were specific to the blue spectral region. Photothermal mechanisms are associated with changes in temperature,(*31*) which we confirmed to be constant throughout our experiments. Finally, photochemical mechanisms are typically driven by stimulation of photosensitive molecules that subsequently drive a downstream chemical reaction. Therefore, our results support the role of a photochemical mechanism of motility in murine MPCs.

As emphasized by Hore(*32*), considerable care must be taken before identifying a particular manifestation of spin-based magnetodetection with a radical pair mechanism. In the case of avian navigation, the photoactive molecule is strongly believed to be Cryptochrome.(*2, 3*) However, in our studies, knockdown of Cry 1 or Cry 2 in MPCs did not affect responsiveness to light. This aligns with previous findings that, *in vitro*, Cry 1 and 2 have a low binding affinity for flavin adenine dinucleotide (FAD), the primary chromophore implicated in the Cryptochrome radical pair mechanism.(*33, 34*) We then performed multi-omics analyses and found that blue light triggers activation of calmodulin proteins. A direct role of calmodulin in blue-light sensing has not been previously identified. However, recent research in diverse species has identified non-canonical photoreceptors, such as LITE-1, Gr28bD, and TRPA1, which couple photon absorption and reactive oxygen species (ROS) to mediate light-avoidance responses.(*35-37*) Notably, calmodulin interacts with transient receptor potential channels to mediate both light and ROS sensing.(*37*) Light-induced ROS generation has also been identified as a putative mechanism linking flavin photochemistry to regulation of ion channels to modulate magnetic field-dependent effects in drosophila.(*38*) Therefore, our data raise the intriguing new hypothesis that calmodulin may participate in quantum coherent radical pair reactions that mediate MPC migration through non-canonical photoreceptors.

Our work has limitations worth noting. Although we did not notice any obvious changes in cell morphology and viability after chlorophyll b treatment, it remains unclear whether chlorophyll b alters other cell behaviors that can affect cell migration. Cells were kept at room temperature during photon recordings, which could have influenced the degree of photon emission observed. Additionally, the relationships identified with multi-omics analyses have yet to be validated. Further investigation into the interactions between light and magnetic fields in the context of muscle injuries could provide valuable insights into the underlying quantum mechanisms that regulate tissue regeneration. These findings have the potential to uncover new pathways for directing regenerative processes, with implications for improving therapeutic strategies.

**Summary and future directions:**

This paper presented evidence that a light-activated, electron spin-based mechanism for quantum magnetodetection guides muscle progenitor cell motility. Further investigation into the interactions between light and magnetic fields in the context of muscle injuries could provide valuable insights into the underlying quantum mechanisms that regulate tissue regeneration. In particular, a comprehensive investigation of the oscillatory field frequencies that disrupt cellular magnetosensitivity could shed light on the identity of the photoactive molecules that initiate electron spin mediated magnetodetection. In addition to potentially extending the existence of quantum compasses from birds to individual cells, the findings presented here the potential to uncover new pathways for directing regenerative processes, with implications for one day improving therapeutic strategies.

**Acknowledgements:**

The author would like to thank Dr. Peter Hore for constructive criticism of the manuscript. The authors would also like to thank Drs. Michael Hatridge and Thomas Purdy for the valuable conversations related to this paper as well as Petra Erdmann-Gilmore, Yiling Mi, Rose Connors, and Dr. Reid Townsend from the Washington University Proteomics Shared Resource (WU-PSR) for performing mass spectrometry proteomics. WU-PSR is supported in part by the WU Institute of Clinical and Translational Sciences (NCATS UL1 TR000448), the Mass Spectrometry Research Resource (NIGMS P41 GM103422) and the Siteman Comprehensive Cancer Center Support Grant (NCI P30 CA091842).

**Funding:**

Research was sponsored by the Army Research Laboratory and was accomplished under Cooperative Agreement Number W911NF-21-2-0208. The views and conclusions contained in this document are those of the authors and should not be interpreted as representing the official policies, either expressed or implied, of the Army Research Office or the U.S. Government. The U.S. Government is authorized to reproduce and distribute reprints for Government purposes notwithstanding any copyright notation herein.

**Author contributions:**

All authors made substantial contributions in the following areas: (1) conception and design of the study, acquisition of data, analysis and interpretation of data, drafting of the article; (2) final approval of the article version to be submitted; and (3) agreement to be personally accountable for the author's own contributions and to ensure that questions related to the accuracy are appropriately investigated, resolved, and the resolution documented in the literature.

The specific contributions of the authors are as follows:

>Conceptualization: FA
>Methodology: GG, KW, MCA, HI, FA
>Software and equipment: HI, BM
>Formal Analysis: KW, GG, MCA, KW, HI
>Investigation: KW, GG, MCA, AU, JB, AW, MM
>Resources: FA
>Data Curation: GG, KW, HI
>Visualization: KW, GG, MCA, HI
>Funding acquisition: DW, DP, SS, FA
>Project administration: KW, GG, FA
>Supervision: FA
>Writing – original draft: KW, GG, MCA, HI, SL, FA
>Writing – review & editing: KW, GG, MCA, HI, JB, AW, BM, MM, DHW, BZ, PC, AU, DP, SS, SL, FA

**Competing interests:** The authors declare that they have no competing interests.

# References

1. Y. Zhang, G. P. Berman, S. Kais, The radical pair mechanism and the avian chemical compass: Quantum coherence and entanglement. *International Journal of Quantum Chemistry* **115**, 1327-1341 (2015).
2. G. Wan, A. N. Hayden, S. E. Iiams, C. Merlin, Cryptochrome 1 mediates light-dependent inclination magnetosensing in monarch butterflies. *Nature Communications* **12**, 771 (2021).
3. J. Xu *et al.*, Magnetic sensitivity of cryptochrome 4 from a migratory songbird. *Nature* **594**, 535-540 (2021).
4. V. Gladkikh, A. Burshtein, G. Angulo, G. Grampp, Quantum yields of singlet and triplet recombination products of singlet radical ion pairs. *Physical Chemistry Chemical Physics* **5**, 2581-2588 (2003).
5. C. T. Rodgers, P. J. Hore, Chemical magnetoreception in birds: the radical pair mechanism. *Proc Natl Acad Sci U S A* **106**, 353-360 (2009).
6. H. M. Blau, B. D. Cosgrove, A. T. Ho, The central role of muscle stem cells in regenerative failure with aging. *Nature medicine* **21**, 854-862 (2015).
7. D. Coletti *et al.*, Static magnetic fields enhance skeletal muscle differentiation in vitro by improving myoblast alignment. *Cytometry Part A: The Journal of the International Society for Analytical Cytology* **71**, 846-856 (2007).
8. J. Stern-Straeter *et al.*, Impact of static magnetic fields on human myoblast cell cultures. *International journal of molecular medicine* **28**, 907-917 (2011).
9. W.-C. Mo *et al.*, Shielding of the geomagnetic field alters actin assembly and inhibits cell motility in human neuroblastoma cells. *Scientific Reports* **6**, 22624 (2016).
10. C. C. Liang, A. Y. Park, J. L. Guan, In vitro scratch assay: a convenient and inexpensive method for analysis of cell migration in vitro. *Nat Protoc* **2**, 329-333 (2007).
11. T. Ritz *et al.*, Magnetic compass of birds is based on a molecule with optimal directional sensitivity. *Biophysical journal* **96**, 3451-3457 (2009).
12. H. G. Hiscock, D. R. Kattnig, D. E. Manolopoulos, P. Hore, Floquet theory of radical pairs in radiofrequency magnetic fields. *The Journal of Chemical Physics* **145**, (2016).
13. K. Maeda *et al.*, Chemical compass model of avian magnetoreception. *Nature* **453**, 387-390 (2008).
14. A. V. Van Huizen *et al.*, Weak magnetic fields alter stem cell–mediated growth. *Science advances* **5**, eaau7201 (2019).
15. K. Wang *et al.*, Bioengineered 3D Skeletal Muscle Model Reveals Complement 4b as a Cell-Autonomous Mechanism of Impaired Regeneration with Aging. *Adv Mater*, e2207443 (2023).
16. S. Suzuki *et al.*, Two-dimensional imaging and counting of ultraweak emission patterns from injured plant seedlings. *Journal of Photochemistry and Photobiology B: Biology* **9**, 211-217 (1991).
17. S. Kwon, W. Yang, D. Moon, K. S. Kim, Biomarkers to quantify cell migration characteristics. *Cancer cell international* **20**, 1-13 (2020).
18. M. Liedvogel *et al.*, Chemical magnetoreception: bird cryptochrome 1a is excited by blue light and forms long-lived radical-pairs. *PLoS One* **2**, e1106 (2007).
19. S. N. Nangle *et al.*, Molecular assembly of the period-cryptochrome circadian transcriptional repressor complex. *Elife* **3**, e03674 (2014).
20. H. Kodama, H. Iijima, Y. Matsui, Systematic identification of exercise-induced anti-aging processes involving intron retention. *bioRxiv*, 2024.2004. 2025.591048 (2024).
21. M. W. Berchtold, A. Villalobo, The many faces of calmodulin in cell proliferation, programmed cell death, autophagy, and cancer. *Biochim Biophys Acta* **1843**, 398-435 (2014).

22. A. Valdeolivas *et al.*, Random walk with restart on multiplex and heterogeneous biological networks. *Bioinformatics* **35**, 497-505 (2019).
23. D. Adler, Z. Shapira, S. Weiss, A. Shainberg, A. Katz, Weak Electromagnetic Fields Accelerate Fusion of Myoblasts. *International journal of molecular sciences* **22**, 4407 (2021).
24. M. Cifra, P. Pospisil, Ultra-weak photon emission from biological samples: definition, mechanisms, properties, detection and applications. *J Photochem Photobiol B* **139**, 2-10 (2014).
25. J. Du *et al.*, The application and trend of ultra-weak photon emission in biology and medicine. *Front Chem* **11**, 1140128 (2023).
26. A. Prasad, A. Balukova, P. Pospisil, Triplet Excited Carbonyls and Singlet Oxygen Formation During Oxidative Radical Reaction in Skin. *Frontiers in physiology* **9**, 1109 (2018).
27. C. M. Mano *et al.*, Excited singlet molecular O(2)((1)Deltag) is generated enzymatically from excited carbonyls in the dark. *Sci Rep* **4**, 5938 (2014).
28. P. N. Youssef, N. Sheibani, D. M. Albert, Retinal light toxicity. *Eye (Lond)* **25**, 1-14 (2011).
29. R. S. Molday, O. L. Moritz, Photoreceptors at a glance. *J Cell Sci* **128**, 4039-4045 (2015).
30. J. X. Tao, W. C. Zhou, X. G. Zhu, Mitochondria as Potential Targets and Initiators of the Blue Light Hazard to the Retina. *Oxid Med Cell Longev* **2019**, 6435364 (2019).
31. R. D. Glickman, Phototoxicity to the retina: mechanisms of damage. *Int J Toxicol* **21**, 473-490 (2002).
32. P. J. Hore, Spin chemistry in living systems. *National Science Review*, nwae126 (2024).
33. N. Karki, S. Vergish, B. D. Zoltowski, Cryptochromes: Photochemical and structural insight into magnetoreception. *Protein Science* **30**, 1521-1534 (2021).
34. R. J. Kutta, N. Archipowa, L. O. Johannissen, A. R. Jones, N. S. Scrutton, Vertebrate cryptochromes are vestigial flavoproteins. *Scientific reports* **7**, 44906 (2017).
35. J. Gong *et al.*, The C. elegans taste receptor homolog LITE-1 is a photoreceptor. *Cell* **167**, 1252-1263. e1210 (2016).
36. C. Montell, Gustatory receptors: not just for good taste. *Current Biology* **23**, R929-R932 (2013).
37. E. J. Du *et al.*, Nucleophile sensitivity of Drosophila TRPA1 underlies light-induced feeding deterrence. *Elife* **5**, e18425 (2016).
38. A. A. Bradlaugh *et al.*, Essential elements of radical pair magnetosensitivity in Drosophila. *Nature* **615**, 111-116 (2023).

# Electron spin dynamics guide cell motility

# Supplementary Materials

**Authors**: Kai Wang,[1-3]* Gabrielle Gilmer,[1,2,4,5]* Matheus Cândia Araña,[1,2,6] Hirotaka Iijima,[1-3] Juliana Bergmann,[1,2] Antonio Woollard,[7] Boris Mesits,[7] Meghan McGraw,[1,2] Brian Zoltowski,[8] Paola Cappellaro,[9,10] Alex Ungar,[9,10] David Pekker,[7] David H. Waldeck,[11] Sunil Saxena,[11] Seth Lloyd,[12] Fabrisia Ambrosio[1-3]**

**Affiliations**:

[1]Discovery Center for Musculoskeletal Recovery, Schoen Adams Research Institute at Spaulding, Boston, MA
[2]Department of Physical Medicine & Rehabilitation, Spaulding Rehabilitation Hospital, Boston, MA
[3]Department of Physical Medicine & Rehabilitation, Harvard Medical School, Boston, MA
[4]Cellular and Molecular Pathology Graduate Program, University of Pittsburgh, Pittsburgh, PA
[5]Medical Scientist Training Program, School of Medicine, University of Pittsburgh, Pittsburgh, PA
[6]École Polytechnique, Institut Polytechnique de Paris, Palaiseau, France
[7]Department of Physics & Astronomy, University of Pittsburgh, Pittsburgh, PA
[8]Department of Chemistry, Southern Methodist University, Dallas, TX
[9]Department of Nuclear Science and Engineering, Massachusetts Institute of Technology, Cambridge, Massachusetts 02139, United States
[10]Department of Physics, Massachusetts Institute of Technology, Cambridge, Massachusetts 02139, United States
[11]Department of Chemistry, University of Pittsburgh, Pittsburgh, PA
[12]Department of Mechanical Engineering, Massachusetts Institute of Technology, Boston, MA
* Equal contribution

**Corresponding Author:

Fabrisia Ambrosio, PhD, MPT

Suite 5.303, 149 13th St, Charlestown, MA 02129

Phone: (412) 657-1525

Email: fambrosio@mgh.harvard.edu

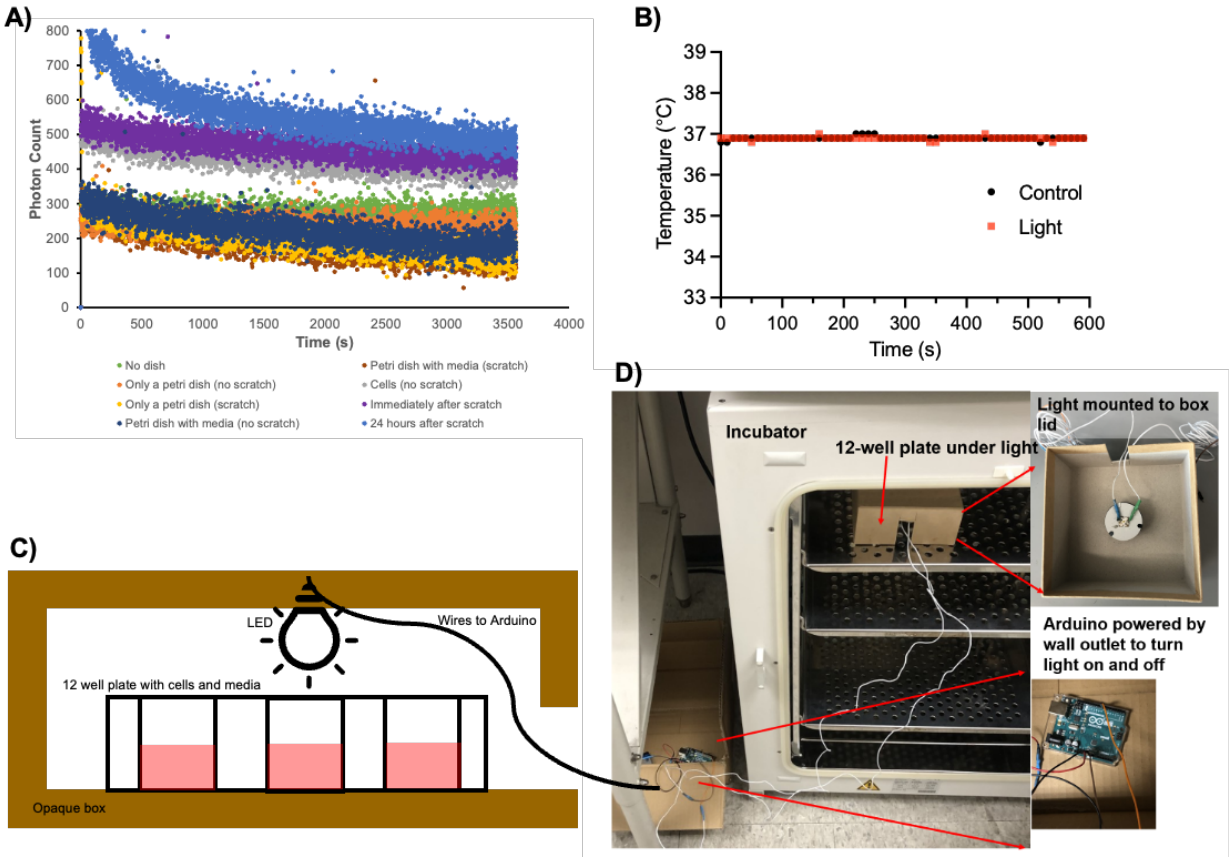

**Figure S1.** A) Raw photon counts emitted from cells, a petri dish, a petri dish with growth media, and no dish under scratch or no scratch conditions. (B) The temperature of cell culture media without (Control) or with (Light) light exposure. The media was exposed to white light for ten minutes at the maximum intensity. (C, D) LED setup for light stimulation experiments.

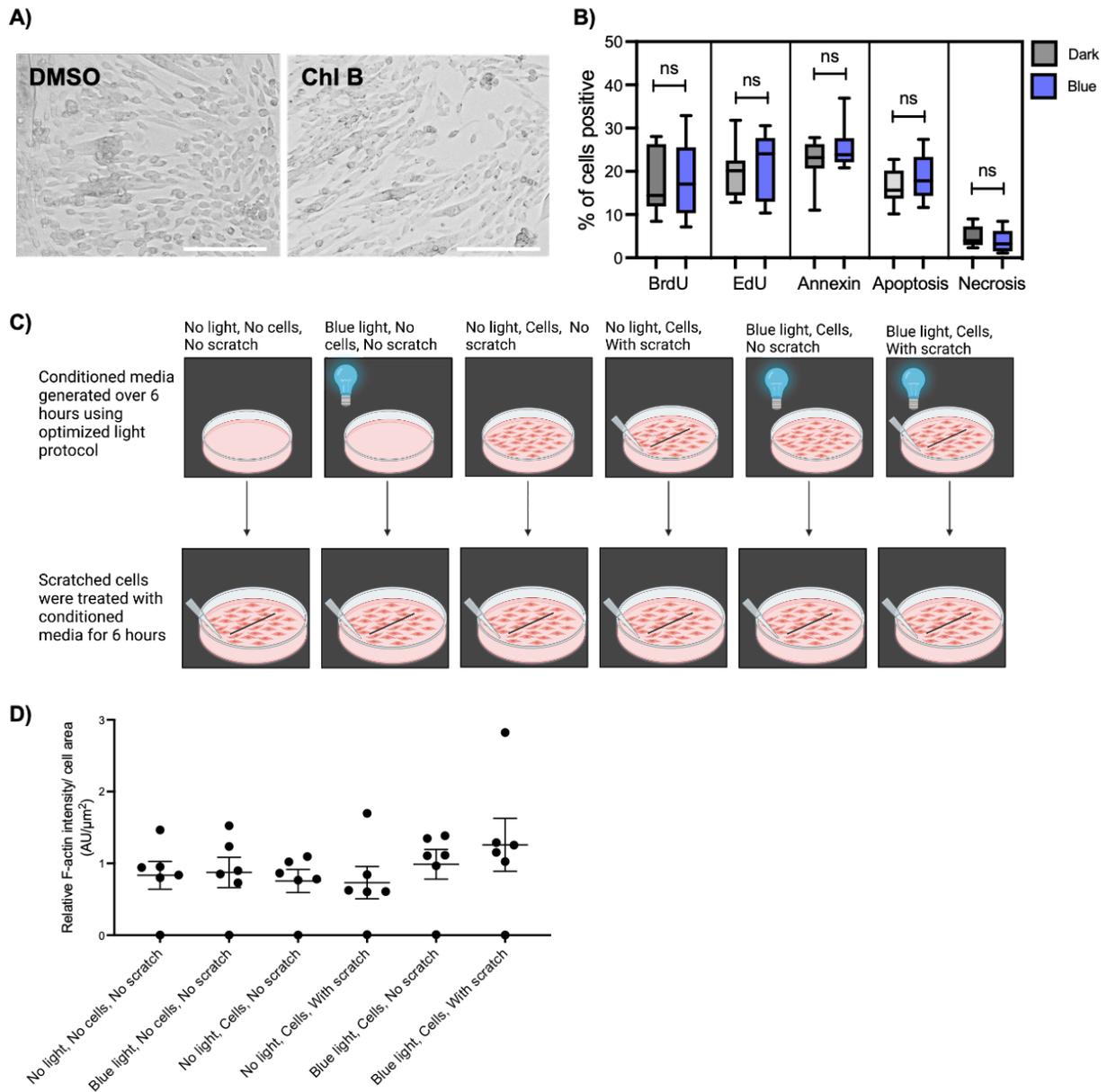

**Figure S2.** (A) Representative images of MPCs after 6 hours of DMSO or Chlorophyll B/DMSO solution . (B) Quantification of cells that are positive for EdU (proliferation), BrdU (proliferation), Annexin V (apoptosis), phosphatidylserine (apoptosis), or membrane-impermeable 7-AAD (necrosis). ns: not significant; Independent samples t-test (EdU and 7-AAD), Mann Whitney U test (BrdU and Annexin V). n=12/group. (C) Experiment design to evaluate the effect of light-conditioned media on cell migration. Created in BioRender. (D) Quantification of F-actin intensity in migrating cells cultured in light-conditioned media. No statistical difference was detected across groups. One-way ANOVA, n = 12/group.

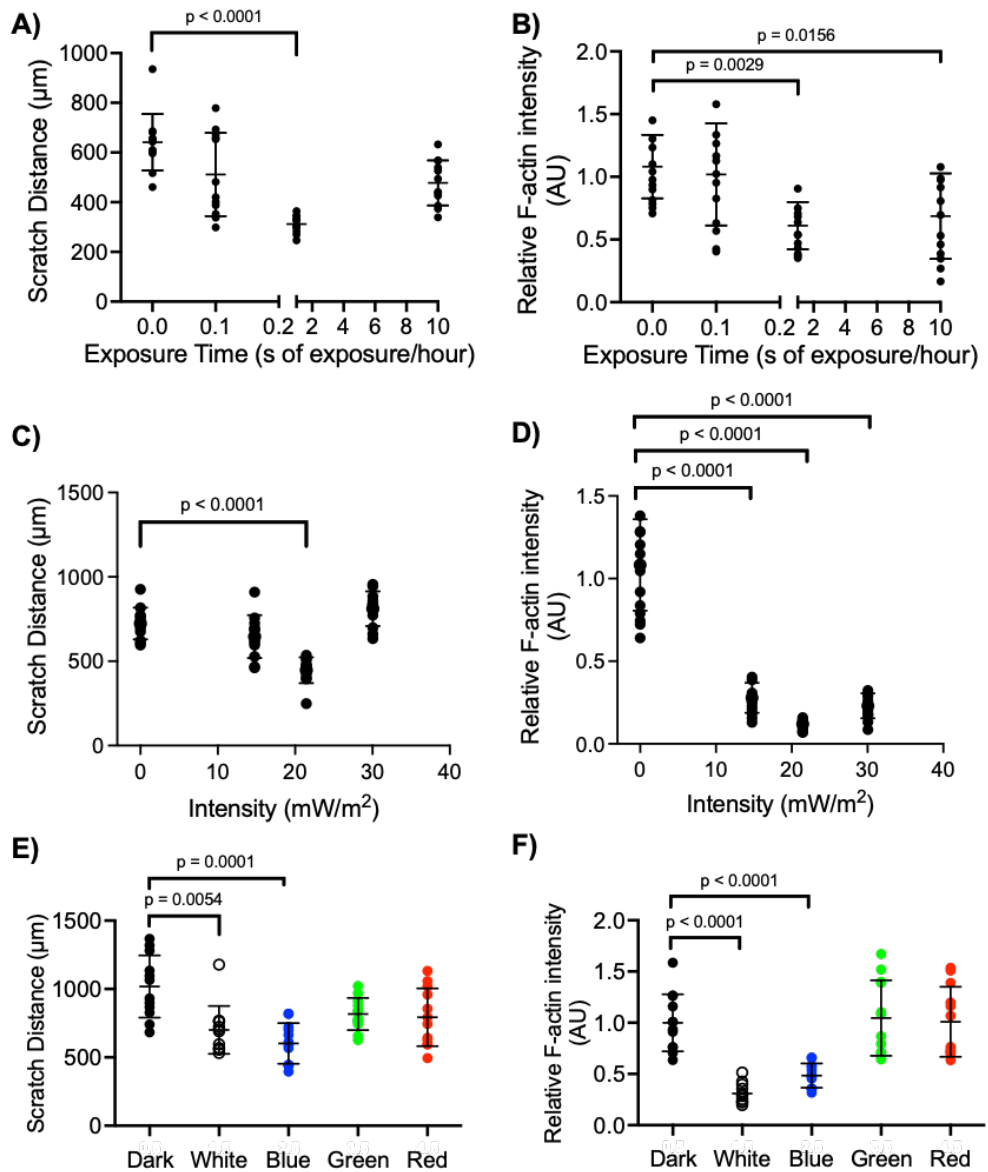

**Figure S3: The effect of light stimulation on female muscle stem cell migration and wound healing.** (A-B) Scratch distance and relative F-actin intensity of scratched female cells after 6 hours of light exposure at various light exposure times. Kruskal Wallis tests, n=12/group. (C-D) Scratch distance and relative F-actin intensity of scratched cells after 6 hours of light exposure at different intensities. Kruskal Wallis tests, n=12/group. (E-F) Scratch distance and relative F-actin intensity versus light spectra. One-way ANOVA, n=12/group.

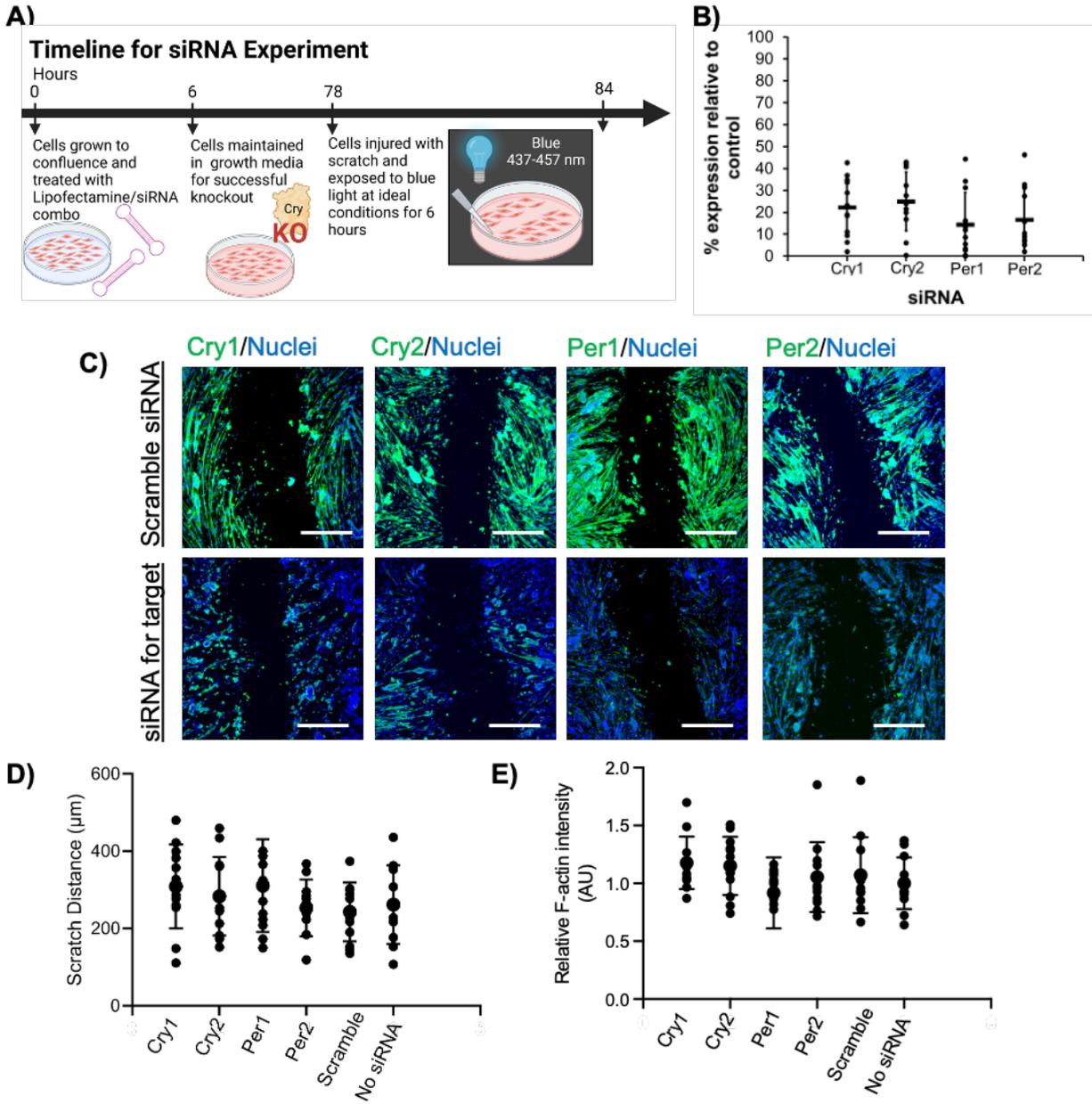

**Figure S4. The effect of cryptochrome and period knockout on cell migration and wound healing.** (A) Timeline for siRNA treatment and light stimulation. (B-C) Quantification of protein intensity and representative images of cells after siRNA treatment. Cry1: Cryptochrome 1; Cry2: Cryptochrome 2; Per1: Period 1; Per2: Period 2. Protein fluorescence intensities after siRNA treatment were normalized to their scramble RNA treatment controls. n = 12/group. Scale bar = 100 μm. (D-E) Scratch distance and relative F-actin intensity of cells after siRNA treatment. No statistical difference was detected across groups. One-way ANOVA, n=12/group.

**Table S1: Equipment and reagent list**

| Equipment/Reagents | Vendor | Catalog # |
|---|---|---|
| C2C12 cells | ATCC | CRL-1772 |
| Andor iXon Ultra 897 EMCCD | Oxford Instruments | DU-897U-CS0-#BV |
| Geltrex | Gibco | A14132-02 |
| PBS | Thermo Scientific | J67653.K2 |
| C57BL/6J mice | Jackson Laboratory | #:000664 |
| HI Horse Serum | Gibco | 26050-088 |
| Hank's Balanced Salt Solution | Gibco | 14175-095 |
| Penicillin-Streptomycin | Sigma Aldrich | P4333-100 mL |
| Collagenase Type II | Worthington | LS004177 |
| Dispase | Gibco | 17105-041 |
| HI FBS | Gibco | 10082-147 |
| Chick embryo extract | US Biological | C3999 |
| Dulbecco's Modified Eagle's Medium - high glucose | Sigma Aldrich | D5796-500mL |
| Basic fibroblast growth factor | PeproTech | 100-18B-50UG |
| 0.5% Trypsin EDTA | Gibco | 15400-054 |
| Royal-Blue (448nm) Rebel LED on a SinkPAD-II 20mm Star Base - 1030 mW @ 700mA | Luxeon Star LEDS | SP-01-V4 |
| Green LUXEON Z on a Saber Z1 10mm Square Base - 92 lm | Luxeon Star LEDS | SZ-01-H1 |
| Far Red (720nm) Rebel LED on a SinkPAD-II 20mm Tri-Star | Luxeon Star LEDS | SP-03-D4 |
| 6500K LUXEON Z White LED on a Saber Z1 10mm Square | Luxeon Star LEDS | SZ-01-K7 |
| ARDUINO UNO REV3 | Arduino | 7630049200050 |
| USB 2.0 CABLE TYPE A/B | Arduino | 7630049201149 |
| Wire PRT-08865 | DigiKeys electronics | 1568-1566-ND |
| Wire PRT-08866 | DigiKeys electronics | 1568-1567-ND |
| Ø1" Bandpass Filter, CWL = 650 ± 8 nm, FWHM = 40 ± 8 nm | ThorLabs | FB650-40 |
| Ø1" Bandpass Filter, CWL = 550 ± 8 nm, FWHM = 40 ± 8 nm | ThorLabs | FB550-40 |
| Ø1" Bandpass Filter, CWL = 450 ± 8 nm, FWHM = 40 ± 8 nm | ThorLabs | FB450-40 |
| Magnaflux Digital White Light / Visible Light Meter | NDT Supplies | 622338 |

| Item | Source | Catalog # |
|---|---|---|
| 10K Ohm Trim Potentiometer Breadboard Trim Potentiometer Kit with Knob Variable Resistors Trimmer Potentiometer Assortment Kit Compatible with Arduino, Blue | Amazon | B09G9TBY38 |
| Paraformaldehyde Solution | Thermo Scientific | J19943-K2 |
| Triton X 100 | Fluka Analytical | 93420-1L |
| Bovine Serum Albumin | Fisher Scientific | BP1600-100 |
| Hoechst 3342, trihydrochloride, trihydrate | Thermo Scientific | H3570 |
| Rhodamine Phallodin | Thermo Scientific | R415 |
| ImageJ | https://ij.imjoy.io/ | |
| Silencer™ Negative Control No. 1 siRNA | Thermo Scientific | 4404021 |
| Per1 siRNA | Thermo Scientific | 4390816 |
| Per2 siRNA | Thermo Scientific | 4390771 |
| Cry1 Mouse RNA silencer | Thermo Scientific | 4390816 |
| Cry2 Mouse RNA silencer | Thermo Scientific | 4390771 |
| BrdU Monoclonal Antibody (MoBU-1) | Thermo Scientific | B35130 |
| Annexin V Polyclonal Antibody | Thermo Scientific | PA5-57231 |
| Apoptosis/ Necrosis Assay Kit (blue, red, green) | Abcam | ab176750 |
| Opti-MEM™ I Reduced Serum Medium | Thermo Scientific | 31985070 |
| Lipofectamine 3000 Transfection Kit | Thermo Scientific | L3000-001 |
| PER1 Polyclonal Antibody | Thermo Scientific | 13463-1-AP |
| PER2 Polyclonal Antibody | Thermo Scientific | PA5-100107 |
| MyoD Antibody | Santa Cruz | sc-377460 |
| Pax7 Antibody | Developmental Studies Hybridoma Bank (DSHB) | Registry ID: AB_528428 |
| Cryptochrome 1 Polyclonal Antibody | Thermo Scientific | 13474-1-AP |
| CRY2 Polyclonal Antibody | Thermo Scientific | PA5-13125 |
| Alexa Fluor 488 goat anti rabbit IgG | Thermo Scientific | A11034 |
| Goat serum | MP Biomedical | 191356 |
| Rneasy Mini Kit (250) | Qiagen | 74106 |
| Sodium dodecyl sulfate | Thermo Scientific | 28312 |
| Ultra Pure 1 M Tris-HCl ph 8 | Thermo Scientific | 15568-025 |
| DTT (dithiothreitol) | Thermo Scientific | R0861 |
| Gikfun DS18B20 Temperature Sensor Waterproof Digital Thermal Probe Sensor for Arduino (Pack of 5pcs) EK1083 | Amazon | B012C597T0 |
| Breadboard Solderless with Jumper Cables– ALLUS BB-018 3Pc 400 Pin | Amazon | 4330587759 |

| Prototype PCB Board and 3Pc Dupont Jumper Wires | | |
|---|---|---|
| Microcon-30kDa Centrifugal Filter Unit with Ultracel-30 membrane | Millipore, | MRCF0R030 |
| Pierce™ Alkylating Reagents | Thermo Scientific | A39271 |
| Sep-Pak tC18 3 cc Vac Cartridge, 200 mg Sorbent per Cartridge, 37 - 55 µm, 50/pk | Waters | WAT054925 |
| Pierce™ Quantitative Peptide Assays & Standards | Thermo Scientific | 23290 |
| Ethylene glycol diethyl ether, 99%, Thermo Scientific Chemicals | Thermo Scientific | 90110 |
| MuMETAL® Zero Gauss Chamber | Magnetic Shield Corp | ZG-206W |
| Chlorophyll b | Sigma Aldrich | 00538 |

**Supplementary Code: Arduino code for light stimulation**

```
int n = 0;
void setup() {
  pinMode(8,OUTPUT);
}
void loop() {
  delay(180);
  while (n <6) {
  digitalWrite(8,HIGH);
  delay(1000); //time of exposure
  digitalWrite(8,LOW);
  delay(3599000); //time of delay (off)
  n = n+1;}
}
```

## Methods

*Materials Availability*

There were no new unique reagents used in this study. All reagents and materials used in these experiments are listed in **Table S1**.

*Data and Code Availability*

All data used to support the conclusions of this study are available from the corresponding author upon reasonable request. The RNA sequencing data is available under the GEO Submission GSE264709. Proteomics data are available on "Massive.ucsd.edu" under the log in "MSV000093662_reviewer" and password "Gabby-006-121323". To download proteomics data, go to, ftp://MSV000093662@massive.ucsd.edu.

*Photon emission experiment set up*

To minimize photon noise, extraneous light was removed from the room using opaque and reflective substances (e.g., construction paper and aluminum foil). All surrounding lights from adjacent rooms and hallways were kept off, and all experiments were conducted between 4:30-6:30 am EST to minimize light from nearby windows. Of note, the photon emission experiments with no bandpass filter were performed in Pittsburgh, while the bandpass filter photon emission experiments were performed in Boston due to lab relocation.

A Zeiss Z01 Observer Microscope or Nikon AXR Confocal was used for photon emission recordings. The magnification was set to 2.5X, the immersion was air, the working distance was 26 mm, and the numerical aperture was 0.55. An Andor iXon Ultra 897 camera was used to record photon emission. The camera was kept off until experiment use and was allowed to cool to -60°C prior to any recordings. Andor Soilis was set to record "Photon Count with Long Exposure (> 10 sec)". Specifically, the exposure time was 0.287 s, the kinetic series length was 1 hour, the shift speed was 0.5 µs, the vertical clock voltage amplitude was "Normal", the readout rate was 1 MHz at 16 bit, the preamplifier gain was set to "Gain 3", and the output amplifier was set to "Electron Multiplying", with the gain level set to 1000. These settings were chosen due to manufacture recommendation of optimal photo collection. Analyses were performed on the middle 500-second window to minimize ambient noise from the lights being on at the beginning of the experiment.

C2C12 myoblasts were plated in Geltrex-coated 35-mm petri dishes. For Geltrex coating, Geltrex was diluted in cold PBS at a 1:100 ratio after thawing on ice, followed by adding to pre-cooled dishes and incubating for 1 hour at 4°C. Geltrex solution was then aspirated, and the dishes were placed in a cell culture incubator (37°C) for 30 minutes. After that, any remaining solution was further aspirated before cell seeding.

C2C12 cells cultured in dishes were kept in the dark for at least 24 hours prior to the experiments, and all photon emission experiments and setup were performed in the dark. After the plate was set over the microscope, a scratch wound was made (see details below). Recordings took place for 1 hour, at which point cells were fixed and stained with Hoechst for cell counts (see details below). Each day, a pair of injured and uninjured cells were recorded to control for batch effects, and the order of recording was randomized.

*Animal husbandry and housing*

MPCs were isolated from young (4-6 months) male and female C57/BL6J mice. Mice were housed in temperature (22°-23°C), humidity (55-65%), and light (12-hour light/dark cycle) controlled environments. Domes were placed in cages for enrichment, food and water access was ad libitum, and 2-4 mice were housed per cage. If any adverse health outcomes, such as pain, lethargy, poor grooming, 20% weight loss, or hunching, were noticed, the mouse was not used for cell isolations, and a veterinarian was consulted immediately to determine an appropriate plan. Animal usage was approved by the University of Pittsburgh's and Massachusetts General Hospital's Institutional Animal Care and Use Committees.

*Cell isolations*

MPC isolations were performed as outlined previously.(*1*) Briefly, on the day prior to isolating cells from the mice, a T75 flask was coated with collagen (1:10 ratio in MilliQ water) and incubated at room temperature overnight. On the day of the cell isolation, this plate was washed with phosphate buffered solution (PBS) twice and allowed to dry at room temperature. Mice were euthanized via $CO_2$ asphyxiation followed by cervical dislocation. Mice were kept on ice prior to dissections, and all cell isolations took place in a biohazard cell culture hood using an aseptic technique. Approximately 5-10 mL of ice-cold wash media (10% horse serum, 1% Penicillin-Streptomycin in Hanks' Balanced Salt Solution (HBSS)) was placed into a sterile petri dish. Sterile tweezers and scissors were used to carefully peel the skin back to expose the limb muscles. Hindlimb and forelimbs were removed from the body and placed in wash media. Each limb was dissected to remove muscle tissue, with care taken to eliminate contamination from the hair, tendons, bone, ligaments, and fat. After all muscle had been removed from the hindlimbs and forelimbs, muscle was chopped using a sterile razor blade to mince muscle into a slurry. The minced muscle was pulled through a 10 mL pipette and placed in a 50 mL tube. Samples were centrifuged at 900g for 5 minutes at 4°C. After centrifugation, the supernatant was removed, and muscle was weighed. Collagenase II (750 U/mL in wash media) was added to the muscle pellet at the following ratio:

$$Amount\ of\ Collagenase\ II\ Solution\ (mL) = \frac{Muscle\ weight\ (g)}{1.6} x10$$

Minced muscle was incubated in Collagenase II solution for 1 hour at 37°C on a shaker. After incubation, cold wash media was added to the tubes to raise the solution volume to 45 mL. Samples were then centrifuged at 900g for 5 minutes at 4°C. The solution was then aspirated back to its original volume. Collagenase II (1000 U/mL) and Dispase (11 U/mL in wash media without horse serum) were added to the solution at a ratio of 1:10 each. The muscle pellet was then pipetted up and down 10 times to thoroughly mix the solution and incubated at 37°C on a shaker for another 40 minutes.

After incubation, ice cold wash media was again added to the solution to raise the supernatant to 45 mL, and the solution was spun at 900g for 5 minutes at 4°C. The supernatant was aspirated, and the muscle pellet was resuspended in 10 mL of ice cold wash media. A 10 mL sterile syringe was connected to a 20 ½ G needle, and the cell solution was plunged up and down through the needle 10 times, with clogging dissipated by wiping the needle with a Kimwipe. Samples were then passed through a 40 µm cell strainer. Wash media (10 mL) was used to wash the original container and also used to pass through and wash the cell strainer. Filtered samples were then centrifuged at 500g for 5 minutes at 4°C.

After centrifugation, the supernatant was removed, and cell pellets were resuspended in 10 mL of MPC growth media (Sterile filtered 20% fetal bovine serum (FBS), 1% Penicillin Streptomycin, 0.05% Chick Embryo Extract in high glucose DMEM). Media with cells were placed on the collagen-coated T75 flask and incubated at 37°C for 24 hours. The next day, Geltrex-coated plates were prepared as previously described. Media and non-attached cells from the collagen-coated plate were then transferred to the Geltrex-coated T175 flasks. Cells were allowed to grow until 70-80% confluence was reached, with media changed every 2-3 days. The same media was used for culturing the cells in all experiments. However, for the experiments involving the Mu-metal Faraday cage and the oscillating magnetic field, 10 ng/mL of basic fibroblast growth factor (bFGF) was added to the media during cell culture.

*Scratch wound assays*

MPCs were cultured in Geltrex-coated 12-well plates and 24-well plates. The scratch wound assay was employed to model cellular injury and assess the healing process. Scratch wounds were generated either manually by an investigator using a sterile 200 µL pipette tip down the center of each well or using a BioTek AutoScratch Wound Making Tool. We note that the manual scratch creates larger wounds than the AutoScratch Tool. This assay induces acute disruption to the myotube cell membrane, with the subsequent "healing" process defined by the rate of wound closure. For the RNA-sequencing and mass spectrometry scratch wound, a more extensive 3 x 3 scratch injury was generated. Wound gap distance and area were measured using ImageJ. After the scratch was made, the MPC media was removed from the wells, and fresh media was added to each well. After scratches were made, cells were then maintained in the cell culture incubators under minimal ambient illumination.

For the light stimulation experiments, 1 mL of MPC media was placed in each well (12-well plates), with $1 \times 10^6$ cells seeded per well. MPCs were grown on 12-well plates until

70-80% confluency was reached and were kept in the dark for 24 hours prior to experiments to minimize the potential impact of autofluorescence. Light-emitting diodes (LEDs) were soldered to a cable set and mounted on the top of an opaque box (**Figure S1C**). Arduinos were used to control light exposure in a pulsed fashion, and the original code used is shown in **Supplementary Code**. The wavelength of light was controlled by taping bandpass filters over the white LED. The intensity of light was measured using a digital visible light meter and adjusted using potentiometers. To control for possible circadian rhythm-based effects and reduce ambient light noise, all experiments were conducted between 4:30-6:30 am EST. The entire setup, including the injury and light exposure procedures, was carried out in a dark room to further minimize external light interference.

For the Faraday cage experiments, 1 mL of media was placed in each well of the 24-well plates, with $0.15 \times 10^6$ MPCs seeded per well. MPCs were grown on 24-well plates for three days prior to experiments starting. The BioTek AutoScratch Wound Making Tool was used to create scratches in each well. Plates were then randomized into two groups: a control group and a group placed inside the Mu-metal Faraday cage for a duration of 6 hours. The Mu-metal Faraday cage used was the MuMETAL® Zero Gauss Chamber from Magnetic Shield Corp., made from high-permeability Mu-Metal to ensure a stable, low-field test environment by attenuating external static and low-frequency AC fields by up to 1,000,000 times. Residual fields of less than 5 nT and low noise levels (<30 fT/sqrtHz) were achieved through degaussing procedures. The cage was placed inside the incubator for at least 24 hours prior to the experiment to ensure thermal equilibrium. The cage is equipped with a hole in the three-layer shield, allowing cables for instruments (such as those for light exposure experiments) to pass through, while also ensuring proper $CO_2$ and temperature control inside the cage.

Chlorophyll B were dissolved in DMSO at the concentration of 1 mM. For chlorophyll B experiment, $0.5 \times 10^6$ cells seeded in each well of 24-well plates and cultured overnight. Scratch wounds were then generated using the AutoScratch Tool. Cells were then washed with growth media twice, followed by culturing in the growth media containing 10 μM chlorophyll B for 6 hours. In the control group, cells were cultured in the growth media containing 10 μL DMSO for 6 hours. The experiments were conducted in a dark room to minimize external light interference.

*Radiofrequency magnetic field experiment*

To evaluate the effects of radiofrequency magnetic fields on MPC migration, custom polydimethylsiloxane (PDMS) cell chambers with glass bottoms were fabricated. For this, Sylgard 184 silicone base was mixed with curing agent in a 10:1 ratio by weight. This mixture was then placed under vacuum to remove any entrapped air. The PDMS mixture was cured by baking the chambers at 100°C for 30 minutes. Cured PDMS rings with an inner diameter of 10 mm were produced and subsequently adhered to glass bottoms using uncured PDMS as an adhesive. The assembled chambers were then baked at 100°C for 30 minutes. Once the chambers were prepared, they were sterilized with 70% ethanol for 30 minutes in a sterile hood, followed by UV light exposure for 30 minutes. Following sterilization, the glass bottoms were coated with Geltrex using the same ratios and steps outlined before for 2 hours

The total volume of each chamber was approximately 300 µL, and 200 µL of media containing $0.3 \times 10^6$ cells/mL were added to each chamber after coating. Cells were allowed to adhere and grow overnight before conducting the experiments. Next, the chambers were placed in a 24-well plate, in which a scratch wound was created using the BioTek AutoScratch wound making tool. The initial wound area was imaged to establish a baseline for subsequent measurements. Following the scratch, the experimental chambers (excluding controls) were placed in the center of a custom-designed coil to generate radiofrequency magnetic fields. The frequency of the oscillating magnetic field was controlled using a RIGOL DG900 high-resolution arbitrary waveform generator. A continuous sinusoidal wave with a 5 Vpp amplitude was applied. Cells were exposed to oscillating magnetic fields at either 1.35 MHz or 11.2 MHz for a duration of 6 hours. Control chambers were only exposed to the GMF during this period. After the 6-hour stimulation, the wound area was imaged again to assess wound healing. The percentage of wound healing was calculated by dividing the wound area after 6 hours of stimulation by the initial wound area measured immediately after scratch.

*Muscle construct preparation and injury*

Muscle constructs were prepared using the protocol we established previously.(*2*) Briefly, primary MPCs isolated from 3 to 5 month old male mice were suspended in a Matrigel/fibrin hydrogel mixture and added to a sterilized polydimethylsiloxane (PDMS) muscle construct frames. Muscle constructs were cultured in proliferation medium (without bFGF) containing 2 mg/ml 6-aminocaproic acid (ACA; Sigma-Aldrich) for 4 days, followed by culturing in differentiation media supplemented with 2 mg/ml ACA for another 14 days to induce myotube formation. To induce muscle construct injury, muscle constructs were incubated with cardiotoxin (0.4 µM, Sigma-Aldrich) for 5 h on a shaker at 37 °C, followed by 20-minute washing with differentiation media for twice. The contractile forces of muscle constructs were measured and analyzed as previously described.(*2*) To assess muscle stem cell lineage progression during regeneration and magnetic field stimulation, muscle constructs were further stained for Pax7 and MyoD, as we previsouly described.(*2*)

*Magnetic field stimulation*

A Helmholtz coils-based magnetic field stimulation chamber was developed to generate static magnetic fields across a multi-well plate. The field strength was measured using a Hall sensor that provided continuous feedback to a Raspberry Pi to maintain the set field or using a Physics toolbox app. For 2D cell monolayer experiments, 1 mL of media was added to each well of the 24-well plates, with $0.15 \times 10^6$ MPCs seeded per well. MPCs were grown on 24-well plates for three days prior to the start of the experiments. The BioTek AutoScratch Wound Making Tool was used to create scratches in each well. Plates were then randomized and placed in either cell culture incubator or the magnetic field stimulation chamber for 6 hours. The cell alignment angle was analyzed using

ImageJ. To apply magnetic field stimulation to muscle constructs, a PDMS insert was used to confine the injured muscle constructs in the direction that myotubes were parallel to the direction of applied magnetic fields. Control groups were the samples in cell culture incubator under the influence of the Earth's magnetic field.

*Immunofluorescence staining and imaging*

All staining took place at room temperature. All plates from each experiment (i.e., data shown on the same figure) were stained at the exact same time. Immediately upon completion of either light exposure or photo emission counting, media was removed, and cells were fixed in 2% paraformaldehyde (PFA) for 10 minutes. Cells were then washed in PBS three times for 2 minutes each and permeabilized with 0.1% triton-X for 15 minutes, followed by washing in PBS three times for 2 minutes each. Cells were then blocked in blocking buffer (0.1% Triton-X + 3% bovine serum albumin in PBS) for 1 hour, after which cells were incubated in rhodamine phalloidin (1:100 dilution in blocking buffer) in the dark for 20 minutes. Cells were then washed with PBS three times for 2 minutes each and stained with Hoechst (1:2,000 in PBS) for 2 minutes. After another three times of PBS wash, cells were maintained in PBS, and plates were wrapped in foil and stores at 4°C prior to imaging.

Imaging took place 1-7 days after staining was complete. Images were taken at 10X magnification on a wide-field Nikon microscope. Four images were taken per well, at the "top", "middle-top", "middle-bottom", and "bottom" of the well. Imaging and exposure parameters (exposure time, gain, LUTs, etc.) were kept constant within each experiment (i.e., data shown on the same figure had the exact same imaging configurations). While imaging, the investigator would measure the scratch distance of each image as the visualized largest distance between the two edges of the cells.

*Image analysis*

Images were exported as TIFF files, and F-actin intensity per unit area was quantified using ImageJ. Prior to analysis, the scale was set to reflect the image dimensions, and image type was changed to RGB color. Measurements were set to include Area and Integrated Density. For each image, 10 randomly selected cells that were close to the edge of the scratch were outlined, and area and integrated density were measured. Care was taken to not include cells that appeared over saturated. F-actin intensity per area was calculated by dividing the raw integrated density by the area. The same investigator performed all image analyses associated with each experiment. The 10 cells were averaged per image, and the four images were averaged to give the F-actin intensity per area values per well that are shown in the figures.

*Silencing RNA treatment*

Silencing RNA (siRNA) were acquired for Period 1 (Per1), Period 2 (Per2), Cryptochrome 1 (Cry1), and Cryptochrome 2 (Cry2), and Lipfectamine 3000 was used to transfect cells with siRNAs. We first performed dosing experiments to determine the appropriate dose of siRNA and lipofectamine for successful transfection. Briefly, MPCs were cultured to 70-80% confluence. We investigated four different combinations for each siRNA: (1) no siRNAs + 1.5 µL lipofectamine, (2) 2 µg of siRNAs + 3 µL lipofectamine, (3) 2 µg of siRNAs + 1.5 µL lipofectamine, and (4) 4 µg of siRNAs + 3 µL lipofectamine.

In 0.5 mL Eppendorf tubes, 100 µL of OptiMEM media was mixed with the above various quantities of lipofectamine. In separate 0.5 mL Eppendorf tubes, the siRNA was mixed with 100 µL of OptiMEM media and 6 µL of Lipfectamine P3000. Both reagents were allowed to incubate for 15 minutes at room temperature. After incubation, media was removed from cells, and 500 µL of warmed DMEM were added to each well (24 well plate used for this experiment). Lipofectamine and siRNA/P3000 mixtures were added to each well and incubated for 6 hours at 37°C. After this incubation, DMEM mixture was removed, and warmed growth media was added to the plates.

Cells were allowed to grow for 72 hours after transfection and then fixed using 2% PFA. All staining steps were completed at room temperature unless otherwise noted. After three 2-minute PBS washes, cells were permeabilized with 0.1% triton-X for 15 minutes. Cells were again washed with PBS three times for 2 minutes each and then blocked with blocking buffer for 1 hour. Primary antibodies for each protein were suspended in blocking buffer + 5% goat serum at the following concentrations: Per1 (1:20), Per2 (1:20), Cry1 (1:20), and Cry2 (1:60). After blocking buffer was removed, primary antibody solution was placed in wells and incubated overnight at 4°C. After overnight incubation, cells were washed with PBS for 2 minutes three times. In the dark, Alexa Fluor 488 anti-rabbit was suspended in blocking buffer + 5% goat serum at a concentration of 1:400, and cells were incubated in the secondary antibody solution for 1 hour. After removal of the secondary, the cells were washed and stained with Hoechst as outlined above after the rhodamine phalloidin stain.

For Per1 and Per2, the 2 µg of siRNAs + 3 µL lipofectamine combination achieved the best protein knockout, while for Cry1 and Cry2, the optimal combination was 4 µg of siRNAs + 3 µL lipofectamine. These concentrations were used in light exposure experiments using the procedures outlined above.

*RNA sequencing*

To increase the surface area of the injury and maximize our ability to detect differences between groups, the injury was modeled by a 3x3 scratch wound across the well. Blue light (437-457 nm) exposure was performed at an intensity of 21 mW/m$^2$ for 1 second of exposure/hour for a total of 6 hours.

RNA was isolated using the Qiagen RNeasy Mini Kit, according to the manufacturer's protocol. Six samples were collected per condition. All collection steps were performed with RNase free pipet tips and tubes. Briefly, cells were detached from plates using trypsin as described above. Cells were suspended in 350 µL of buffer RLT and homogenized on ice using a tissue grinder. 350 µL of 70% ethanol was mixed with the cell solution. 700 μL of sample solution were transferred to a RNeasy Mini spin column and spun for 15 seconds at 10,000 g. Supernatant was removed, and 700 μL of Buffer RW1 was added to the column. Sample was again spun at 10,000g for 15 seconds, and the supernatant was discarded. 500 μL of Buffer RPE was added to the column, which was then spun at 10,000g for 15 seconds, and this step was repeated. The spin column was then placed in a new 1.5 mL collection tube, and 50 μL of RNase-free water was added to the column. Samples and columns were spun for 1 minute at 10,000g and stored at -80°C until shipment to Novogene.

Bulk RNA-sequencing was performed by Novogene. A priori, it was determined that samples would be excluded if the Phred score was below 20. For the analyses presented here, all samples scored a Phred score above 30. RNA sequencing analyses were performed in Partek Flow workspace. Samples were aligned to the STAR 2.7.10b mus musculus (mm10) reference genome. Initial data processing was performed as outlined by Partek Flow bulk RNA-sequencing tutorial.

*Mass spectrometry proteomics*

MPCs were plated, injured, and exposed to light as described above for RNA sequencing. Six samples were collected. After trypsinization, cell pellets were resuspended in 60 μL of sodium dodecyl sulfate (SDS) buffer (4% SDS in 100 mM Tris-HCl, pH 8.5; UA buffer) and sonicated using a tissue homogenizer. Samples were then reduced with 100 mM DTT at 95°C for 10 minutes and then frozen at -80°C until shipment to the Washington University at St. Louis Proteomics core.

Protein extraction at the proteomics core was performed using well-established protocols.(3, 4) Briefly, peptides were transferred to the top of a 30,000 molecular weight cut-off filter and spun in a microcentrifuge (Eppendorf) at 10,000 rcf for 10 minutes. An additional 300 μl of UA buffer was added and the filter was spun at 10,000 rcf for 10 minutes in a microcentrifuge. The flow through was discarded and the proteins were alkylated using 100 μl of 50 mM Iodoacetamide (IAM) in UA buffer. IAM in UA buffer was added to the top chamber of the filtration unit. The samples were gyrated at 550 rpm using a Thermomixer (Eppendorf) at room temperature for 30 minutes in the dark. The filter was spun at 10,000 rcf for 10 minutes and the flow through discarded. Unreacted IAM was washed through the filter with two additions of 200 μl of UA buffer, and centrifugation at 10,000 rcf for 10 minutes after each buffer addition. The UA buffer was exchanged with digestion buffer (DB), (50 mM ammonium bicarbonate buffer). Two sequential additions of DB (200 μl) with centrifugation after each addition to the top chamber was performed. The filters were transferred to a new collection tube and samples were digested with a combination of LysC (1 mAU per filter) and trypsin (1:50 wt/wt) in DB buffer on top of the filter for two hours and overnight at 37 °C. The filters were spun at 14,000 rcf for 15 minutes to collect the peptides in the flow through. The filter was washed with 50 μl 100mM ammonium bicarbonate buffer and the wash was collected with the peptides. In preparation for desalting, peptides were acidified to 1% (vol/vol) TFA final concentration.

Peptides were desalted using SepPak. The peptides were eluted from the SepPak in 50% acetonitrile (MeCN), 0.1% trifluoroacetic acid (TFA into 1.5 ml tubes and lyophilized. The peptides were dissolved in 100 µl of 1% MeCN in water. An aliquot (10 %) was removed for quantification using the Pierce Quantitative Fluorometric Peptide Assay kit. 1 µg total peptide from each sample was transferred to an AS vial for label-free analysis (LFQ), the remaining peptides were lyophilized and stored at -80ºC for TMT labeling.

The lyophilized peptides (50 µg) from 6 bioreplicates of each condition were dissolved in 20 µl of HEPES buffer (100 mM, pH 8.5) and labeled according to the vendor protocol using the TMT-10 reagent kit. The labeled samples were combined into groups of eight samples and two reference pools (15 µg peptide from each of the samples in the study combined), dried, and dissolved in 120 µl of 1 % (vol/vol) formic acid (FA). The three combined TMT-10 labeled samples were desalted using a SepPak as described above.

Offline basic reverse phase fractionation was performed as previously described.[102] The HPLC system was prepared by purging solvent lines A and B with their respective buffers. The flow rate for equilibration was 1 ml/min. The column was equilibrated with 100% of Solvent A (4.5 mM ammonium formate pH 10, 2% MeCN). The gradient method for basic pH reversed-phase chromatography at a flow rate of 1 ml/min was as follows (time in min.), %B (4.5 mM ammonium formate pH 10, 90% MeCN): 0, 0; 7, 0; 13, 16; 73, 40; 77, 44; 82, 60; 98, 60; 100, 0; 120, 0. Collected fractions were concatenated as previously described[102] and each fraction was transferred to AS vials for global data analysis.

The unlabeled peptides were analyzed using trapped ion mobility time-of-flight mass spectrometry.(5) Peptides were separated using a *nano-ELUTE → chromatograph* (Bruker Daltonics. Bremen, Germany) interfaced to a timsTOF Pro mass spectrometer (Bruker Daltonics) with a modified nano-electrospray source (CaptiveSpray, Bruker Daltonics). The mass spectrometer was operated in PASEF mode.(5) The samples in 2 µl of 1% (vol/vol) FA were injected onto a 75 µm i.d. × 25 cm Aurora Series column with a CSI emitter (Ionopticks). The column temperature was set to 50 °C. The column was equilibrated using constant pressure (800 bar) with 8 column volumes of solvent A (0.1% (vol/vol) FA). Sample loading was performed at constant pressure (800 bar) at a volume of 1 sample pick-up volume plus 2 µl. The peptides were eluted using one column separation mode with a flow rate of 300 nL/min and using solvents A (0.1% (vol/vol) FA) and B (0.1% (vol/vol) FA/MeCN): solvent A containing 2%B increased to 17% B over 60 min, to 25% B over 30 min, to 37% B over 10 min, to 80% B over 10 min and constant 80% B for 10 min. The MS1 and MS2 spectra were recorded from *m/z* 100 to 1700.

The collision energy was ramped stepwise as a function of increasing ion mobility: 52 eV for 0–19% of the ramp time; 47 eV from 19–38%; 42 eV from 38–57%; 37 eV from 57–76%; and 32 eV for the remainder. The TIMS elution voltage was calibrated linearly using the Agilent ESI-L Tuning Mix (*m/z* 622, 922, 1222).

The labeled peptides were analyzed using high-resolution nano-liquid chromatography tandem mass spectrometry (LC-MS). Chromatography was performed with an Acclaim

PepMap 1000 C18 RSLC column (75 μm i.d. × 50 cm; Thermo-Fisher Scientific) on an EASY-*nano*LC 1000 (Thermo Fisher Scientific). The column was equilibrated with 11 μl of solvent A (1% (vol/vol) formic acid (FA)) at 700 bar pressure. The samples in 1% (vol/vol) FA were loaded (2.5 μl) onto the column with 1% (vol/vol) FA at 700 bar. Peptide chromatography was initiated with mobile phase A (1% FA) containing 5% solvent B (100 % ACN, 1 % FA) for 5 min, then increased to 23% B over 100 min, to 35% B over 20 min, to 95% B over 1 min and held at 95% B for 19 min, with a flow rate of 250 nl/min. Data were acquired in data-dependent mode. Full-scan mass spectra were acquired with the Orbitrap mass analyzer using a scan range of *m/z* = 350 to 1500 and a mass resolving power set to 70,000. Twelve data-dependent high-energy collisional dissociations were performed with a mass resolving power at 35,000, a fixed lower value of *m/z* 100, an isolation width of 1.2 Da, and a normalized collision energy setting of 32. The maximum injection time was 60 ms for parent-ion analysis and 120 ms for product-ion analysis. Ions that were selected for MS/MS were dynamically excluded for 40 sec. The automatic gain control (AGC) was set at a target value of 3e6 ions for full MS scans and 1e5 ions for MS2.

For timsTOF files, data from the mass spectrometer were converted to peak lists using DataAnalysis (version 5.2, Bruker Daltonics). The machine data from the LC-MS analysis of isobarically-labeled peptides, using the Q-Exactive mass spectrometer, were converted to peak lists using Proteome Discoverer (version 2.1.0.81, ThermoScientific). MS2 spectra with parent ion charge states of +2, +3 and +4 were analyzed using Mascot software[104] (Matrix Science, London, UK; version 2.8.0.1) against a concatenated UniProt (ver January 2023) database of mouse (17,264 entries) and common contaminant proteins (cRAP, version 1.0 Jan. 1st, 2012; 116 entries). Trypsin/P enzyme specificity with a maximum of 4 missed cleavages allowed was used. The searches were performed with a fragment ion mass tolerance of 20 ppm and a parent ion tolerance of 20 ppm for Q Exactive™ data. Label-free "single-shot" LC-MS data from the timsTOF mass spectrometer were searched with a fragment ion mass tolerance of 40 ppm and a parent ion tolerance of 20 ppm. Carbamidomethylation of cysteine was specified in Mascot as a fixed modification. Deamidation of asparagine, formation of pyro-glutamic acid from N-terminal glutamine, acetylation of protein N-terminus, oxidation of methionine, and pyro-carbamidomethylation of N-terminal cysteine were specified as variable modifications.

*Differential gene expression analysis*

Data preprocessing and differential gene expression analyses were performed in accordance with the established workflow suggested by Law with cut-off thresholds of log$_2$ fold change >1.5 and false discovery rate <0.05.(6) Raw count data were normalized by counts per million (CPM). We removed genes that have very low counts (<11) in RNA-seq data prior to downstream analysis on biological and statistical grounds(7). From a biological point of view, a gene must be expressed at some minimal level before it is likely to be translated into a protein or to be considered biologically important. From a statistical point of view, genes with consistently low counts or low signal intensity are very unlikely be assessed as differentially expressed because low counts or low signal intensity do not provide enough statistical evidence for a reliable judgement to be made. Such genes were therefore removed from the analysis without any loss of information.(6) We used the filterByExpr function of the edgeR R package with default parameters for low count data (min count = 10).(8) The count data were further normalized using Trimmed Mean Mvalue (TMM) normalization and voom transformation in the edgeR and limma R package.(8, 9) The differential gene expression analysis was performed using the limma R package.(10) The Benjamini–Hochberg false discovery rate (FDR) control for multiple hypothesis testing was used to produce q-values.

*Functional characterization of genes or proteins*

To map transcriptomic and protein responses together, log$_2$ fold change (light, injury / no light, injury) values across the RNA-seq and mass-spectrometry proteomics data was merged using the R function intersect. Biological function of genes or proteins of interests were determined using enrichr.(11) The Benjamini-Hochberg FDR control for multiple hypothesis testing was used to produce q-values. REVIGO software(12) was used to summarize redundant GO terms and the results were visualized as a treemap using the R function treemap.

*Directional gene set enrichment analysis*

Traditional gene set enrichment analysis (GSEA) uses weighted gene ranks based on the association with diseases of interest.(13) However, this conceptual framework is not limited to diseases. GSEA can also be used to evaluate the association with specific gene expression patterns.(14) Directional GSEA (dGSEA) is an extension of GSEA that uses gene ranks based on the association of specific gene set (source genes) and the other genes (target genes) to comprehensively assess biological functions or signaling pathways associated with the specific gene set. The weights of gene ranks are the sums of Pearson's correlation coefficient across the source gene in each target gene. dGSEA was performed using the R/Bioconductor package genekitr.(15) The Benjamini-Hochberg FDR control for multiple hypothesis testing was used to produce q-values. Leading edge analysis was performed after dGSEA to determine the core genes defining the subset of genes with positive contribution to the enrichment score before it reaches its peak.That is, those that are most correlated with the phenotype of interest.(13)

*Network propagation using Random Walk Regression (RWR)*

We first constructed protein interactive network by accessing protein network data from String database (version 12.0, full network, scored links between proteins).(16) For the protein interactive network, RWR was performed by R/Bioconductor package

RandomWalkRestartMH(17) with Calmodulin family proteins (Camk2a, Camk2b, Camk2d, Camk2g, Camkk1, and Calm1) proteins used as seed nodes, inspired by the CytoTalk algorithm.(18) RWR simulates a walker starting from one node or a set of nodes (seed nodes) in one network, and such walker randomly moves in the network to deliver probabilities on the seed nodes to other nodes. After iteratively reaching stability, the affinity score of all nodes in the given network to seeded node were obtained. Affinity scores (higher score indicates neighbor of Calmodulin family proteins) were used for subsequent GSEA.

*Statistical analyses*

The primary endpoint of this study was distance across the scratch, and we performed an *a priori* power analysis to estimate the needed sample size. For this, a variability analysis on manual scratch distance was performed after an investigator scratched 50 independent wells, and the average scratch width was 790 ± 140 µm. Based on this variability, an average change in scratch distance of 200 µm would be the equivalent of detecting an effect size of 0.58. To detect this effect size with an alpha level of 0.05 and a statistical power of 0.88, 12 samples were needed per group. Our secondary outcome measure was F-actin intensity per unit cell area. A recent paper used live cell imaging to show that F-actin intensity per cell area is correlated with more recent migration and faster cell migration.(19) Thus, F-actin intensity per unit cell area served as a surrogate measure for migration in our experiments. For RNA sequencing and mass spectrometry proteomics, a sample size of 6 was used per condition, as recommended by previous studies.(20, 21)

Statistical analyses were performed in SPSS Statistics, Version 28.0 (IBM Corp., NY, USA) or Prism GraphPad, Version 10.3.1. The data were displayed as means with a standard deviation of the mean (means ± SEM). A Shapiro–Wilk test was initially performed to check the normality of data. When conditions for normality were not met, groups were compared using a Mann–Whitney U test or nonparametric ANOVA with Dunn's multiple comparisons. When normality conditions were met, two-tailed Student's t-test or one-way ANOVA with post-hoc Tukey HSD (honestly significant difference) test was performed for two-group and multiple group comparisons, respectively. Statistical tests used for each individual assessment were also noted in the figure legends. Statistical significance was defined *a priori* as an alpha level of 0.05.

*Rigor & reproducibility*

For all experiments presented in this study, plates were randomized to experimental condition, and the investigator performing the analysis was blinded to experimental groups. To minimize inter-rater reliability, the same investigator performed scratches within each experiment, and the same investigator completed all analyses for that experiment. Plates were not excluded from analyses unless they became contaminated during the experiment. We did not register our experimental protocols *a priori*. This study was conducted in accordance with the ARRIVE Guidelines 2.0.(22)

# References

1. K. Wang *et al.*, Bioengineered 3D Skeletal Muscle Model Reveals Complement 4b as a Cell-Autonomous Mechanism of Impaired Regeneration with Aging. *Adv Mater* **35**, e2207443 (2023).
2. K. Wang *et al.*, Bioengineered 3D Skeletal Muscle Model Reveals Complement 4b as a Cell–Autonomous Mechanism of Impaired Regeneration with Aging. *Advanced Materials* **35**, 2207443 (2023).
3. J. Erde, R. R. Loo, J. A. Loo, Improving Proteome Coverage and Sample Recovery with Enhanced FASP (eFASP) for Quantitative Proteomic Experiments. *Methods Mol Biol* **1550**, 11-18 (2017).
4. P. Mertins *et al.*, Reproducible workflow for multiplexed deep-scale proteome and phosphoproteome analysis of tumor tissues by liquid chromatography-mass spectrometry. *Nat Protoc* **13**, 1632-1661 (2018).
5. F. Meier *et al.*, Online Parallel Accumulation-Serial Fragmentation (PASEF) with a Novel Trapped Ion Mobility Mass Spectrometer. *Mol Cell Proteomics* **17**, 2534-2545 (2018).
6. C. W. Law *et al.*, RNA-seq analysis is easy as 1-2-3 with limma, Glimma and edgeR. *F1000Res* **5**,  (2016).
7. Y. Chen, A. T. Lun, G. K. Smyth, From reads to genes to pathways: differential expression analysis of RNA-Seq experiments using Rsubread and the edgeR quasi-likelihood pipeline. *F1000Res* **5**, 1438 (2016).
8. M. D. Robinson, D. J. McCarthy, G. K. Smyth, edgeR: a Bioconductor package for differential expression analysis of digital gene expression data. *Bioinformatics* **26**, 139-140 (2010).
9. C. W. Law, Y. Chen, W. Shi, G. K. Smyth, voom: Precision weights unlock linear model analysis tools for RNA-seq read counts. *Genome Biol* **15**, R29 (2014).
10. M. E. Ritchie *et al.*, limma powers differential expression analyses for RNA-sequencing and microarray studies. *Nucleic Acids Res* **43**, e47 (2015).
11. M. V. Kuleshov *et al.*, Enrichr: a comprehensive gene set enrichment analysis web server 2016 update. *Nucleic Acids Res* **44**, W90-97 (2016).
12. F. Supek, M. Bošnjak, N. Škunca, T. Šmuc, REVIGO summarizes and visualizes long lists of gene ontology terms. *PLoS One* **6**, e21800 (2011).
13. A. Subramanian *et al.*, Gene set enrichment analysis: a knowledge-based approach for interpreting genome-wide expression profiles. *Proc Natl Acad Sci U S A* **102**, 15545-15550 (2005).
14. J. Lamb *et al.*, A mechanism of cyclin D1 action encoded in the patterns of gene expression in human cancer. *Cell* **114**, 323-334 (2003).

15. Y. Liu, G. Li, Empowering biologists to decode omics data: the Genekitr R package and web server. *BMC Bioinformatics* **24**, 214 (2023).
16. D. Szklarczyk *et al.*, The STRING database in 2023: protein-protein association networks and functional enrichment analyses for any sequenced genome of interest. *Nucleic Acids Res* **51**, D638-d646 (2023).
17. A. Valdeolivas *et al.*, Random walk with restart on multiplex and heterogeneous biological networks. *Bioinformatics* **35**, 497-505 (2019).
18. Y. Hu, T. Peng, L. Gao, K. Tan, CytoTalk: De novo construction of signal transduction networks using single-cell transcriptomic data. *Sci Adv* **7**,  (2021).
19. S. Kwon, W. Yang, D. Moon, K. S. Kim, Biomarkers to quantify cell migration characteristics. *Cancer Cell Int* **20**, 217 (2020).
20. L. Yu, S. Fernandez, G. Brock, Power analysis for RNA-Seq differential expression studies. *BMC Bioinformatics* **18**, 234 (2017).
21. E. S. Nakayasu *et al.*, Tutorial: best practices and considerations for mass-spectrometry-based protein biomarker discovery and validation. *Nat Protoc* **16**, 3737-3760 (2021).
22. N. Percie du Sert *et al.*, The ARRIVE guidelines 2.0: Updated guidelines for reporting animal research. *Br J Pharmacol* **177**, 3617-3624 (2020).